\documentstyle[11pt,psfig]{article}
\begin{document}

\begin{titlepage}
\thispagestyle{empty}
\begin{flushright}
Utrecht-THU-95/26\\
hep-th/9601073\\
\end{flushright}
\vskip 1cm
\begin{center}
{\LARGE\bf  Hawking Spectrum and High Frequency Dispersion}\\
\vskip 1cm
{\large Steven Corley and Ted Jacobson}
\vskip .5cm
{\it Institute for Theoretical Physics, University of Utrecht\\
P.O. Box 80.006, 3508 TA Utrecht, The Netherlands}\\
and\\
{\it Department of Physics, University of Maryland\\
                          College Park, MD 20742-4111, USA}\\
             {\tt corley, jacobson@umdhep.umd.edu}
      
\end{center}
\vskip 1cm

\begin{abstract}

We study the spectrum of created particles in two-dimensional black hole geometries for a linear, hermitian scalar field satisfying a Lorentz non-invariant field equation with higher spatial derivative terms that are suppressed by powers of a fundamental momentum scale $k_0$. The preferred frame is the ``free-fall frame" of the black hole. This model is a variation of Unruh's sonic black hole analogy. We find that there are two qualitatively different types of particle production in this model: a thermal Hawking flux generated by ``mode conversion" at the black hole horizon, and a non-thermal spectrum generated via scattering off the background into negative free-fall frequency modes. This second process has nothing to do with black holes and does not occur for the ordinary wave equation because such modes do not propagate outside the horizon with positive Killing frequency. The horizon component of the radiation is astonishingly close to a perfect thermal spectrum: for the smoothest metric studied, with Hawking temperature $T_H\simeq0.0008k_0$, agreement is of order $(T_H/k_0)^3$ at frequency $\omega=T_H$, and agreement to order $T_H/k_0$ persists out to $\omega/T_H\simeq 45$ where the thermal number flux is $O(10^{-20}$). The flux from scattering dominates at large $\omega$ and becomes many orders of magnitude larger than the horizon component for metrics with a ``kink", i.e. a region of high curvature localized on a static worldline outside the horizon. This non-thermal flux amounts to roughly 10\% of the total luminosity for the kinkier metrics considered. The flux exhibits oscillations as a function of frequency which can be explained by interference between the various contributions to the flux.

\end{abstract}
\end{titlepage}
\vskip 2mm

\section{Introduction} 

Black holes are boost machines. They process high frequency input and 
deliver it as low frequency output, owing to the gravitational redshift.
Thus they provide a glimpse of the world at very short distance scales.
This short distance world consists of nothing but vacuum fluctuations. 
A black hole acts like a microscope, allowing us to peer into the vacuum 
and see something of the nature of these short distance fluctuations.

When looking directly at a black hole, we see the vacuum fluctutations
as processed by quantum field dynamics. In ordinary continuum quantum
field theory, this processing results in Hawking radiation\cite{Hawking}, 
with a perfect
thermal spectrum. The older the black hole, the higher is the boost 
interpolating between the input and output. In fact this boost grows 
exponentially as $\exp(t/4M)$ with the age $t$ of the hole. Therefore,
according to ordinary quantum field theory, the phenomenon of Hawking 
radiation involves physics at arbitrarily high frequencies and short 
distances.
If there is new physics at some length scale, then the output of the black 
hole will be the result of processing at least down to that scale. 

Perhaps, therefore,  the existence and properties of Hawking radiation can 
teach us something about physics at very short distances. Note that 
the term ``short" here refers to measurements in the asymptotic rest
frame, or the free-fall frame, of the black hole. If one assumes exact
Lorentz invariance and locality, 
the large boosts provided by the black hole are just 
symmetry transformations, and one can learn nothing new. However, 
the assumption of unlimited boost invariance
is beyond the range of observational support, 
so we shall not make 
it. Instead, we consider in this paper the effect on black hole radiation of 
(local and non-local) Lorentz non-invariant modifications to quantum field theory.

It is worth pointing out that even if one does assume exact Lorentz
invariance, there is still room for short distances to play a crucial
role in black hole physics. One way this might happen is via nonlocality,
such as in string theory\cite{Lowetal}. Another possibility is that the infalling matter or vacuum fluctuations might 
have intense gravitational interactions with the outgoing
trans-Planckian degrees of freedom\cite{tHooft}. Thus, even
if exact Lorentz invariance is assumed, the use of ordinary field theory 
in analyzing physics around black holes might be unjustified. 

To some extent, one can sidestep the short-distance regime by imposing a
boundary condition on the quantum field in a timelike region outside the
event horizon\cite{Ted2}. Assuming that field modes propagate in the ordinary
way below some cutoff frequency $\omega_c$, and assuming that the 
outgoing modes with frequencies below the cutoff but well above the 
Hawking temperature $T_H$ are in their ground state, then the usual 
Hawking effect can be deduced in an approximation that gets better 
as $T_H/\omega_c$ gets smaller. This calculation shows that a 
conservative upper bound on the 
deviations from the thermal spectrum is of order 
$O((T_H/\omega_c)^{1/2})$. Other estimates\cite{Ted1}, based on the 
behavior of accelerated detectors near the horizon or on the 
trace anomaly, suggest that the deviations will be much smaller, of order 
$O(T_H/\omega_c)$. Such arguments leave some room for interesting
dependence on short distance physics however, due to cumulative effects.

In view of the gentle curvature of spacetime outside the black hole one might
expect no excitation of high energy degrees of freedom. On the other hand, 
even if there is only a small amplitude for excitation in a time of order
$M$, it is conceivable that the amplitude has a secular part which grows
with time as the trans-Planckian degrees of freedom creep, while redshifting
over extremely long times (e.g. $M^3$), away from the horizon. 
(For the purposes of this paper the term ``Planck scale" will refer to the 
scale at which hypothetical Lorentz non-invariant physics occurs.) 
Furthermore,
even small deviations in the spectrum might have a large effect when 
integrated all the way up to trans-Planckian wavevectors.
  
Given a particular model of short distance physics, we would thus like to ask 
the following questions:
\begin{enumerate}
\item Where do the outgoing modes come from?
\item Does the above mentioned out vacuum boundary condition hold?
\item Exactly how large are the deviations from the thermal Hawking flux?
\item Are the deviations from the thermal Hawking flux small even at 
very short wavelengths? 
\item Do the deviations for short wavelengths accumulate to make 
a large difference in any physical quantity, such as the energy flux
or energy density?
\end{enumerate}

The simple model we shall consider in this paper is a quantum field 
in two spacetime dimensions satisfying a linear wave equation with higher spatial derivative terms.
The dispersion relation $\omega=\omega(k)$ thus differs at high wavevectors 
from that of the ordinary wave equation. The particular dispersion
relation we shall study in detail is $\omega^2=k^2-k^4/k_0^2$.
A modified dispersion relation occurs ubiquitously in
all sorts of physical situations. Whenever there is new structure at some 
scale, for example as in a plasma or a crystal,  wave propagation senses this, and the 
structure is reflected in the dispersion relation. Unruh\cite{Unruh2} recently studied a model like this which was motivated by a 
sonic analog of a black hole. 
Although he describes the model in terms of sound propagation in an 
inhomogeneous background fluid flow, the model is in fact identical to that
of a scalar field in a black hole spacetime, with the co-moving frame of the 
background flow replaced by the free-fall frame of the black hole. 
 
By numerical integration of the altered partial differential wave equation (PDE), 
Unruh studied the propagation of wavepackets in this model and 
established that, to the numerical accuracy of his calculation, Hawking radiation still occurs and is unaffected by the altered dispersion relation. 
The numerical accuracy was not quite good enough to rule out deviations at the upper bound referred to above. 
Perhaps the most interesting thing about the model is the peculiar behavior 
of wavepackets sent backwards in time toward the horizon: rather than 
getting squeezed in an unlimited way against the horizon and ceaselessly
blueshifting, the wavepackets reach a minimum distance of approach, then 
reverse direction and propagate back away from the horizon.
The blueshift at the closest approach to the horizon is independent of 
the retarded time about which the outgoing wavepacket was centered, and
the packet continues to blueshift on the way out going backwards in time.

Subsequently, Brout, Massar, Parentani and Spindel (BMPS) \cite{BMPS} made an analytical study of the Unruh model, and came to similar conclusions in a leading order approximation in $1/M$.
In addition, BMPS introduced another model, differing from the Unruh model
in that the altered dispersion relation is defined with respect to 
Eddington-Finkelstein coordinates. In the BMPS model, an outgoing wavepacket
propagated backward in time does not reverse direction but rather  
 hugs the horizon at a distance of one ``Planck length", 
with exponentially growing wavevectors. For this model 
the usual Hawking effect at leading order in $1/M$ was established
by analytical methods.

The primary purpose of the present paper is to determine precisely 
the spectrum of 
Hawking radiation for a model with a nonlinear dispersion relation as in the 
Unruh and BMPS models. In our model the field equation has fourth order
spatial derivatives in the free-fall frame. We wish to evaluate quantitatively the deviations
from the thermal spectrum, including the high wavevector region.
To achieve this aim, numerical integration of the 
PDE is impractical (at least for us), 
and leading order approximations are insufficient. Instead, we employ a two-pronged attack. First,
we exploit the stationarity of the background metric to simplify the problem.
Thus, instead of solving a PDE, we numerically solve alot of ordinary 
differential equations (ODE's) for the mode functions. Second, as a check on 
the accuracy of our numerical solutions, we develop the exact solution 
for a subclass of the models. To our surprise we have found in the models
studied here that, in addition to the Hawking radiation, radiation is produced via scattering from the static curvature.

A second purpose of our paper is to give a physical picture of the Hawking
effect in the context of these models with altered dispersion relation.
What we describe has alot in common with the picture explained by BMPS in 
\cite{BMPS}
(although we developed our picture independently before becoming aware
of their paper). 
The picture has two essential features, reversal of group velocity without
reflection and ``mode conversion" from one branch of the dispersion relation
to another. Interestingly, both these phenomena can occur for linear waves
in inhomogenous plasmas\cite{Skiff,Stix,Swanson}, and undoubtedly occur in many other settings as well.
The propagation of a wavepacket and the direction-reversal phenomenon can
be understood using the WKB approximation. At the turn-around point 
partial mode conversion from a positive 
free-fall frequency to a negative free-fall frequency wave takes place. 
This mode conversion gives rise to the Hawking effect. 

A third purpose of our paper is to discuss  the ``stationarity puzzle"
in these models: 
If the wavepackets go from infinity to infinity, without ever passing through the collapsing matter, then how can there be any particle production? 
  
The remainder of our paper is organized as follows. 
Section 2 defines the model
to be studied and section 3 describes the wavepacket propagation, mode conversion, and scattering in this model using a WKB analysis.  Section 4 lays out the computational techniques we employed to obtain the precise quantitative results that are reported and interpreted in section 5.
In section 6 the stationarity puzzle is discussed, and section 7 
contains a summary of our results. 
Throughout the paper we use units in which $\hbar=c=G=1$, unless otherwise
specified.

\section{The model and its quantization}

The model we shall consider consists of a free, hermitian scalar field propagating in a two dimensional black hole spacetime. The dispersion relation for the field
lacks Lorentz invariance, and is specified in the free-fall frame of the 
black hole, that is, the frame carried in from the rest frame at infinity by 
freely falling trajectories. This is the same frame as the one used in
 the Unruh model \cite{Unruh2}, but the dispersion relation we adopt
is different. The BMPS model on the other hand adopts the same dispersion 
relation as Unruh, but applys it in the Eddington-Finkelstein coordinate 
frame.

\subsection{Field equation}
\begin{figure}[hbt]
\centerline{
\psfig{figure=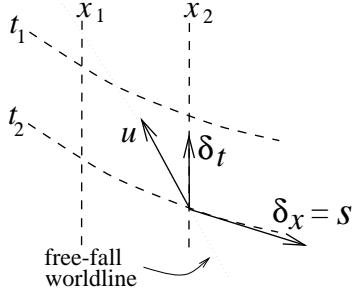,angle=-90,height=4cm}}
\caption{\small A patch of spacetime showing a free-fall trajectory
and some $t$ and $x$ (Lema\^{\i}tre-like) coordinate lines. $u$ and $s$ are orthonormal vectors, and the derivative along $s$ is modified,
while that along $u$ is just the partial derivative. The notations $\delta_t$ and $\delta_x$ denote $\partial/\partial t$
and $\partial/\partial x$ respectively, and $\delta_t$ is the Killing vector.}
\label{lemaitre}
\end{figure}

Let $u^\alpha$ denote the unit vector field tangent to 
the infalling worldlines, and let $s^\alpha$ denote the orthogonal, 
outward pointing, unit vector, so that $g^{\alpha\beta}=u^\alpha u^\beta-s^\alpha s^\beta$.
(See Fig. \ref{lemaitre}.)

The action is assumed to have the form:
\begin{equation}
S={1\over2}\int d^2x \, \sqrt{-g} g^{\alpha\beta} 
{\cal D}_\alpha\phi^* {\cal D}_\beta\phi,
\label{action}
\end{equation}
where the modified differential operator ${\cal D}_\alpha$ is defined by 
\begin{eqnarray}
u^\alpha {\cal D}_\alpha &=& u^\alpha \partial_\alpha\\
s^\alpha {\cal D}_\alpha &= &\hat{F}(s^\alpha \partial_\alpha).
\end{eqnarray}
The time derivatives in the local free-fall frame are thus left unchanged,
but the orthogonal spatial derivatives are replaced by 
$\hat{F}(s^\alpha \partial_\alpha)$.
The function $\hat{F}$ determines the dispersion relation. For the moment it will be left unspecified. 
Invariance of the action (\ref{action}) under 
constant phase transformations of $\phi$ guarantees that there is a conserved
current for solutions and a conserved ``inner product" for pairs of solutions 
to the equations of motion. However, since 
${\cal D}_\alpha$ 
is {\it not} in general a derivation, simple integration by parts is not 
allowed in obtaining the equations of motion or the form of the current. We shall obtain these below after further specifying the model.

The black hole line elements we shall consider are static and have the form
\begin{equation}
ds^2=dt^2-(dx-v(x)\, dt)^2.
\label{metric}
\end{equation}
This is a generalization of the Lema\^{\i}tre line element for the Schwarschild
spacetime, which is given by $v(x)=-\sqrt{2M/x}$ (together with the usual angular part). We shall assume $v<0$, $dv/dx>0$, and 
$v\rightarrow v_o$ as $x\rightarrow\infty$.
$\partial_t$ is a Killing vector, of squared norm 
$1-v^2$, and the event horizon is located at $v=-1$. 
The curves given by $dx-v\, dt=0$ are timelike free-fall worldlines 
which are at rest (tangent to the Killing vector) where $v=0$. 
Since we assume $v<0$ these are {\it ingoing} trajectories. 
$v$ is their coordinate velocity, $t$ measures proper time along them,
and they are everywhere orthogonal to the constant $t$ surfaces
(see Fig. \ref{lemaitre}). We
shall refer to the function $v(x)$ as the {\it free-fall velocity}.
The asymptotically flat region corresponds to $x\rightarrow\infty$.

In terms of the notation above, the orthonormal basis vectors adapted to the free-fall frame
are given by $u=\partial_t+v\partial_x$ and $s=\partial_x$, and 
and in these coordinates $g=-1$. Thus the action (\ref{action}) becomes
\begin{equation}
S={1\over2}\int dtdx\, 
\Bigl(|(\partial_t+v\partial_x)\phi|^2 -|\hat{F}(\partial_x)\phi|^2\Bigr).
\end{equation}
If we further specify that $\hat{F}(\partial_x)$ is an odd function of $\partial_x$,
then ${\cal D}_\alpha$ becomes a derivation ``up to total derivatives",
and integration by parts yields the field equation 
\begin{equation}
(\partial_t+\partial_x v)(\partial_t+v\partial_x)\phi=
\hat{F}^2(\partial_x)\phi.
\label{eom}
\end{equation}
The conserved inner product in this case is given by 
\begin{equation}
 (\phi,\psi) = i \int dx\,  \Bigl(\phi^{*}
(\partial_{t} + v\partial_{x})\psi - \psi(\partial_{t} + v\partial_{x})
\phi^{*}\Bigr),
\label{inner}
\end{equation}
where the integral is over a constant $t$ slice and is independent of $t$  if $\phi$ and $\psi$ satisfy the field equation (\ref{eom}). The inner product can of course be 
evaluated on other slices as well, but it does not take the same simple form 
on other slices\footnote{In fact the inner product is non-local when evaluated on other slices if $\hat{F}(s^\alpha\partial_\alpha)$ is nonlocal. The conserved 
current density $j^\alpha$ is determined by the equation 
$\partial_\alpha j^\alpha=
{\cal D}_a\Bigl(\sqrt{-g}g^{\alpha\beta}
(\phi_1^*{\cal D}_\beta \phi_2-\phi_2{\cal D}_\beta\phi_1^*)\Bigr)$.}.
 
\begin{figure}[h]
\centerline{
\psfig{figure=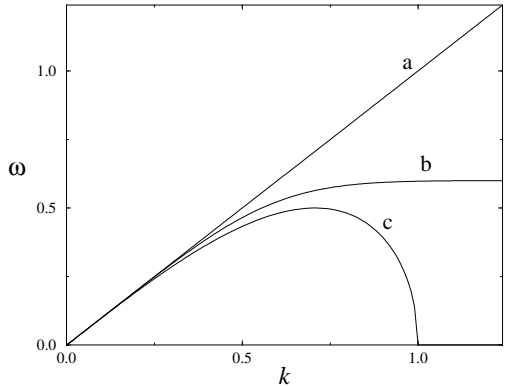,angle=-90,height=6cm}}
\caption{\small  
Curve $a$ is the standard dispersion relation for the massless wave equation,
curve $b$ is the type used by Unruh, and curve $c$ is the one used in 
this paper (\ref{F}).}
\label{F1}
\end{figure}

The dispersion relation for this model in flat spacetime, or in the local
free-fall frame (assuming $v(x) \approx constant$), is given by 
\begin{equation}
\omega^2=F^2(k),
\label{disp}
\end{equation}
where $F(k)\equiv -i\hat{F}(ik)$.
Unruh's choice for the function $F(k)$ has the property that 
 $F^{2}(k)$ = $k^{2}$ for $k \ll k_{0}$ and $F^{2}(k)$ = $k_{0}^2$ for $k \gg k_{0}$, where $k_{0}$ is a wavevector characterizing the scale of the new 
physics. We usually think of $k_0$ as being around the Planck mass. 
Specifically, he considered the functions 
$F(k)=k_0\{\tanh[(k/k_0)^n]\}^{1\over n}$. Of course there are many other modifications one could consider.  Perhaps the simplest is given by
\begin{equation}
F^{2}(k) = k^{2} - k^4/k_{0}^2.
\label{F}
\end{equation}  
This dispersion relation has the same small $k$ behaviour
as Unruh's, but behaves quite differently for large $k$.
It has the technical advantage that the field equation (\ref{eom})
has no derivatives higher than fourth order, and  for this reason
it is the one on which all the calculations in this paper are based, 
although we shall briefly discuss the behavior for alternate choices in the
final section.
These two dispersion relations are plotted in Figure \ref{F1} along with the dispersion relation for the ordinary wave equation. 

\subsection{Quantization}
To quantize the field we assume the field operator $\hat\phi(x)$ is 
self-adjoint
and satisfies the equation of motion (\ref{eom}) and the canonical commutation relations. In setting up the canonical formalism, it is simplest to use the 
time function and evolution vector for which only first order time 
derivatives appear in the action. (Otherwise one must introduce extra momenta
which are constrained, and then pass to the reduced phase space.) 
This just means that we define the momenta by 
$$\pi=\delta L/\delta (\partial_t\phi)=(\partial_t +v\partial_x)\phi,$$
i.e., $\pi$ is the time derivative along the free-fall world lines.
The equal time canonical commutation relations are then 
$[\phi(x),\pi(y)]=i\delta(x,y)$, as usual.

We define an annihilation operator corresponding to an initial data set
$f$ on a surface $\Sigma$ by
\begin{equation}
a(f)=( f,\hat{\phi}),
\label{C2}
\end{equation}
where the inner product is evaluated on $\Sigma$.
If the data $f$ is extended to a solution of the field
equation then we can evaluate the inner product (\ref{C2}) on
whichever surface we wish. The hermitian adjoint of $a(f)$
is called the creation operator for $f$ and it is given by
\begin{equation}
a^{\dagger}(f)=-( f^*,\hat{\phi}).
\label{C2*}
\end{equation}
The commutation relations between these operators follow
from the canonical commutation relations satisfied by the
field operator. The latter are equivalent to
\begin{equation}
 [a(f),a^{\dagger}(g)]=( f, g),
\label{C3}
\end{equation}
provided this holds for all choices of $f$ and $g$.
Now it is clear that only if $f$ has positive, unit norm
are the appelations ``annihilation" and ``creation"
appropriate for these operators.
{}From (\ref{C3}) and the definition of the inner product
it follows identically that we also have the commutation
relations
\begin{equation}
[a(f),a(g)]=-( f, g^*),\qquad
[a^{\dagger}(f),a^{\dagger}(g)]=-( f^*, g).
\end{equation}

A Hilbert space of ``one-particle states" can be defined by
choosing a decomposition of the space $S$ of complex
initial data sets (or solutions to the field equation) into a
direct sum of the form $S=S_p\oplus S_p{}^*$, where all the
data sets in $S_p$ have positive norm and the space $S_p$
is orthogonal to its conjugate $S_p{}^*$. Then all of the
annihilation operators for elements of $S_p$ commute with
each other, as do the creation operators. A ``vacuum" state
$|\Psi\rangle$ corresponding to $S_p$ is defined by the condition
$a(f)|\Psi\rangle=0$ for all $f$ in $S_p$, and a Fock space of
multiparticle states is built up by repeated application
of the creation operators to $|\Psi\rangle$.

It is not necessary to construct a specific Fock space in order to 
study the physics of this system. In fact, any individual positive
norm solution $p$ defines annihilation and creation operators and
a number operator $N(p)=a^\dagger(p)a(p)$. The physical significance
of the number operator depends of course on the nature of $p$. 

There are two types of positive norm wavepackets in which we are
interested. The first are those corresponding to the quanta of Hawking
radiation. These have positive Killing frequency, that is, 
they are sums of solutions satisfying $\partial_t\phi=-i\omega\phi$ 
with $\omega>0$. It is not obvious that such solutions have positive norm 
in the inner product (\ref{inner}), and in fact they do not in general. 
However, using the fact that 
the Killing frequency is conserved, we know that if a positive Killing 
frequency wavepacket were to propagate out to infinity (or any other region where $v=0$),
the integral for its norm would be manifestly positive. Since the 
norm is conserved, this suffices. 

The other type of positive norm wavepackets we shall employ are those
which correspond to particles as defined by the free-fall observers.
These have positive free-fall frequency, that is, they are sums of solutions satisfying $(\partial_t+v\partial_x)\phi=-i\omega'\phi$, with $\omega'>0$, 
on some time slice. These have manifestly positive norm 
(if the solutions summed are orthogonal to each other like, for example, 
harmonic modes in a constant $v$ region), although the free-fall frequency is {\it not} conserved. 
 
Finally, we conclude this section on quantization with a cautionary remark.
One sees from the dispersion relation (\ref{F}) that, for $k^2>k_0^2$,
the field has imaginary frequency modes which are well-behaved in space.
In principle these modes must be included in a complete quantization of 
the model. Although imaginary frequency modes can be 
quantized\cite{Fulling,Kang}, the resulting model is unstable in that the 
energy spectrum is unbounded below. However, these modes play no role
in our analysis of the Hawking effect, so we shall simply ignore them as 
an irrelevant unphysical feature of the model.

\section{Wavepacket propagation and mode conversion}
\label{wpandmc}

In this section we describe, by way of pictures, the production of 
Hawking radiation
from an initial vacuum state by means of a process known as ``mode
conversion".  We also describe a new process of particle production
via scatering in a static geometry that happens in the dispersive
models studied here. We assume that all 
ingoing positive free-fall frequency wavepackets are
unoccupied, at some given time, far (but not infinitely far---see 
section \ref{puzzle}) from the hole where $v(x)$ is approximately 
(or exactly) constant. Given
this initial state, we wish to calculate the number of particles, in
a given outgoing packet, detected by an observer far from the hole who is
at rest with respect to the hole.   
Following the standard technology (see section \ref{pc}), the number of particles in this packet is obtained by propagating the packet back in time to where the initial ground state boundary condition is imposed and taking the norm of its negative free-fall frequency piece.

The behavior of a wavepacket propagated back in time can be understood
qualitatively as follows.  
Assume a solution to the field equation (\ref{eom}) of the form 
$\phi = e^{-i \omega t} f(x)$ and solve the resulting ODE (\ref{ode}) for 
$f(x)$ by the WKB approximation. That is, write $f(x) = \exp(i\int k(x)\,dx)$ 
and assume the quantities $\partial_x v$ and $\partial_x k/k$ are negligible
compared to $k$. 
The resulting equation is the position dependent dispersion relation
\begin{equation}
(\omega - v(x)k)^{2} = F^{2}(k).
\label{disp'}
\end{equation}
This is just the dispersion relation in the local free-fall frame, since the 
free-fall frequency $\omega'$ is related to the Killing frequency $\omega$ by
\begin{equation}
\omega'=\omega-v(x)k.
\label{otrans}
\end{equation}
The position-dependent dispersion relation is useful for understanding
the motion of wavepackets that are somewhat peaked in both position
and wavevector. A graphical method we have employed is described below.
The same method was used by BMPS\cite{BMPS}, who also found a Hamiltonian
formulation for the wavepacket propagation using Hamilton-Jacobi theory.

\begin{figure}[h]
\centerline{
\psfig{figure=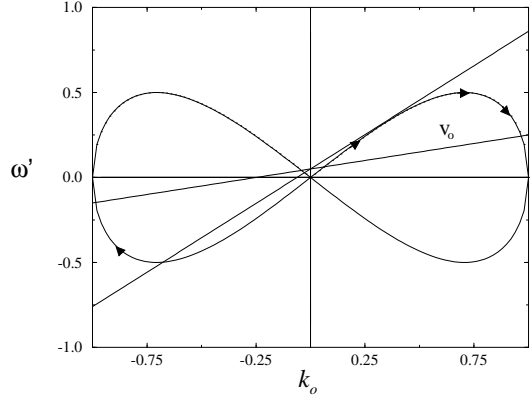,angle=-90,height=6cm}}
\caption{\small Graphical solution of the 
position-dependent dispersion relation 
(\ref{disp'}), with $F(k)$ given by (\ref{F}), in units where $k_0=1$. 
The line labeled $v_{0}$ corresponds to a position far from the hole.  
The other line corresponds to the classical turning point.  The $k$ values of the intersections of the straight and curved lines are the solutions to the dispersion relation for fixed $\omega$ and $v$.  For the $v_{0}$ line these are   denoted from left to right by $k_{-}$, $k_{-s}$, $k_{+s}$, and $k_{+}$ in the text. The filled arrowheads indicate the 
direction of propagation of wavepackets, in momentum space, as discussed in the text.}
\label{dispreln}
\end{figure}
 
Graphs of the square root of both sides of equation (\ref{disp'}) are
shown in Figure \ref{dispreln} for $F(k)$ given by (\ref{F}) and for 
two different values of $v$. 
As $x$ varies, the slope $-v(x)$ ($=|v(x)|$) of the straight line representing
the left hand side of (\ref{disp'}) changes, but for a given wavepacket
the intercept $\omega$ is fixed since the Killing frequency is conserved.
For a given $x$, the
intersection points on the graph correspond to the possible wavevectors
in this approximation. These solutions to the dispersion relation for 
fixed $\omega$ and $v$ will be denoted, in increasing order, as
\begin{equation}
k(\omega)=k_{-},\; k_{-s}, \; k_{+s}, \; k_{+}.
\label{roots}
\end{equation}
(The subscript ``$s$" is intended to suggest ``smaller" in absolute value.)
Note that for the ordinary wave equation one would have only the two
roots with $\omega>0$ corresponding to $k_{-s}$ and $k_{+s}$ at the velocity $v_0$.

The coordinate velocity 
$dx/dt$ of a wavepacket is the group velocity $v_{g} = d\omega/dk$. 
This may also be expressed, using (\ref{otrans}), 
as $v_{g} = v_g' + v(x)$, where $v_{g}' \equiv d\omega '/dk$ is the group velocity in the free-fall frame  
which corresponds to the slope of the curved line in Figure \ref{dispreln}. 
The group velocity is positive at $k_{+s}$ and negative at 
$k_{-}$, $k_{-s}$ and $k_{+}$. Thus while there is one
outgoing mode at fixed positive $\omega$, there are three ingoing
modes. Of crucial importance is the fact that the $k_-$ mode outside 
the horizon ($v>-1$) has {\it negative} free-fall frequency when the 
Killing frequency is positive.

Now consider a 
wavepacket located far from the hole, centered about
frequency $\omega$, and containing only $k$ values around  $k_{+s}$. 
This is an outgoing wavepacket so, going 
backwards in time, the packet moves towards the hole. Two qualitatively
different effects govern the wavepacket propagation, namely, mode conversion
at the horizon and scattering off the geometry. These will now be discussed
in turn.

\subsection{Mode conversion at the horizon}
\label{conversion}
As the wavepacket propagates backwards in time towards the black 
hole $|v(x)|$ increases, so the slope of the straight line in Figure \ref{dispreln} increases, until 
eventually the straight line becomes tangent to the dispersion curve.
At this point $v_{g}$ drops to zero.  If $\omega$ is very small compared to $k_0$, then this stopping point $x_t$ occurs when $v(x)$ is very close to $-1$, that is, just barely outside the horizon. 


What happens at the stopping point? It was incorrectly suggested in 
Ref. \cite{Ted1} that the wavepackets just asymptotically approach limiting position $x_t$ and wavevector $k_t$. 
However, near the stopping point the point particle picture of the wavepacket motion is inadequate, and the spread in both $k$ and $x$ must be considered. 
One can determine qualitatively what happens by considering the behavior of 
nearby solutions to the dispersion relation as follows.\footnote{In fact, the
WKB approximation breaks down as the stopping point is approached, however this
does not prevent us from obtaining qualitative information about the motion
of the wavepacket as described here.}
As pointed out by Unruh\cite{Unruhpc}, it is an unstable situation for the 
wavepacket to just sit at the stopping point: for $k$ slightly 
above $k_t$ the group velocity drops below zero (i.e. the
comoving group velocity drops below the magnitude of the free-fall velocity)
so, backwards in time,  the wavepacket tends to move back {\it away}
from the horizon and therefore to the right (to higher wavevectors) on the dispersion curve. Once this begins to happen, $k$ continues to increase
as the wavepacket moves further away. Exactly this behavior was found in 
Unruh's numerical solution\cite{Unruh2} to the PDE. In brief, a long
wavelength $k_{+s}$-packet went in, and a short wavelength $k_{+}$-packet came out! This is an example of the phenomenon of ``mode conversion"\cite{Stix,Swanson}, but it is only half the story.


There is another short wavelength solution to the
dispersion relation as $x$ approaches $x_t$, at $k_{-}$ on the negative wavevector,
negative free-fall frequency branch of the dispersion curve, that 
mixes in. We will discuss in a moment
a quantitative measure of the relative amplitudes of the $k_{-}$ and 
$k_{+}$ packets arising from this mode conversion process, 
by looking at how it works
for the usual wave equation. Suffice it to say here that 
the negative wavevector mode mixes in strongly for suffiently small $\omega$ for both Unruh's dispersion relation and
(\ref{F}), as shown both by Unruh's solution of the PDE and by the ODE methods applied by BMPS\cite{BMPS} and ourselves.
The ``converted", negative wavevector, wavepacket also has a negative group velocity, and so also moves, backwards in time, away from the hole.
 The end result thus consists of two wavepackets, one constructed of large positive $k$ wavevectors and the other of large negative
$k$ wavevectors, both propagating away from the hole 
(at different group velocities) and reaching
the asymptotically flat (constant free-fall velocity) region.  
The number of created particles in the final, late time, wavepacket is 
given by (minus) the norm of the negative wavevector (and negative free-fall frequency) part of the initial, early time wavepacket.

Let us see how the conversion amplitude is determined in the case of the 
ordinary wave equation with the linear dispersion relation. This will also indicate how it works for the nonlinear dispersion relations.
First note that the wavepacket vanishes inside the horizon (from the causal behavior of the ordinary wave equation), so it must have {\it some} negative wavevector component, since a purely positive wavevector wavepacket cannot vanish on the half line (or any open interval). But how large is this negative wavevector piece?
The WKB form of a single frequency mode is 
$\phi\sim\exp(i\int k\, dx)$, and the dispersion relation $\omega-vk=k$ yields $k=\omega/(1+v)$. Expanding about the horizon at $x=0$
we have $v\simeq -1+\kappa x$, where $\kappa=v'(0)$, 
so $k\simeq \omega/\kappa x$, so 
$\phi\sim\exp\Bigl(i(\omega/\kappa)\ln x\Bigr)$.\footnote{This 
derivation of the form of the mode function can be questioned
on the grounds that the WKB approximation may not be valid. {\it Sufficient} 
conditions for validity of WKB are  $dk/dx\ll k^2$ and $v'\ll k$. 
The dispersion relation implies $dk/dx\simeq(\kappa/\omega)\, k^2$, 
so both of these conditions are satisfied only when $\omega\gg\kappa$.
Nevertheless for some reason (perhaps related to conformal invariance)
the WKB solution is in fact exact for the ordinary wave equation.}
This mode function can be analytically
continued to the upper half $x$-plane, yielding a purely positive wavevector
function $\phi_+$ on the real axis, or to the lower half $x$-plane, yielding a purely negative wavevector function $\phi_-$. These two functions agree for positive values of $x$ but differ for negative values, the difference being given by the discontinuity across the branch cut of the logarithm. Writing 
$-x=xe^{\pm i\pi}$, one sees that the ratio of the values at any negative 
$x$ is given by $\phi_-/\phi_+=\exp(2\pi\omega/\kappa)$. Hence, to obtain a wavepacket that vanishes for negative $x$, one must form the combination 
$\phi_+-\exp(-2\pi\omega/\kappa)\phi_-$. The exponential factor has the form 
$\exp(-\omega/T)$, where $T=\kappa/2\pi$ is the Hawking temperature. 
Thus the negative wavevector part is large only when $\omega$ 
is not too much larger than the Hawking temperature.

Now consider how things change when the altered dispersion relation is
used. The wavepacket can be propagated in fairly close to the horizon 
before the wavevectors reach the nonlinear part of the dispersion curve,
so there will be some domain near the horizon in which it takes approximately the same form as for the ordinary wave equation. The negative 
wavevector part can be inferred in this domain from the above argument, so one
sees that as long as this domain of concurrence exists, the conversion 
amplitude to the negative wavevector branch should be approximately the same.
The essential difference from the ordinary wave equation is just that, rather than remaining crammed against the horizon, the short wavelength wavepackets
propagate back away from the horizon.

It is worth emphasizing that our analysis of the wavepacket motion
assumes that 
the asymptotic velocity is not too small. Otherwise, as the 
negative wavevector wavepacket is propagated backwards in
time, Fig. \ref{dispreln} shows that eventually a point is reached
where the comoving group velocity passes from $- \infty$ to $+ \infty$.
This issue is addressed further in section \ref{puzzle}.

\subsection{Scattering}
\label{scattering}
If the ordinary two-dimensional wave equation were satisfied, there would be
no scattering off the geometry at all, because this equation is conformally invariant and all two-dimensional line elements are conformally flat.
However, the higher spatial derivative terms in (\ref{eom}) spoil conformal
invariance, hence there is some scattering. The reversal of group velocity
at the horizon described in the previous subsection is already an example of 
this, but in addition there will be scattering from the geometry outside the
black hole. For metrics with a minimum characteristic length scale 
$l\gg k_0^{-1}$ this scattering will be extremely small, since the long 
wavelength modes almost satisfy the ordinary wave equation, and the short
wavelength modes do not ``see" the scale $l$.

\begin{figure}[tb]
\centerline{
\psfig{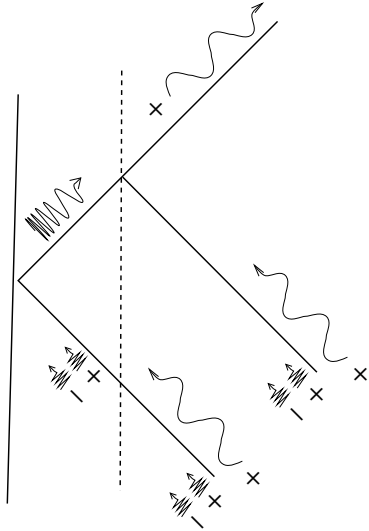}}
\caption{\small Schematic representation of the history of an outgoing
low frequency positive wavevector wavepacket. The solid vertical
line is the horizon, and the dashed line is the ``kink" where the
scattering takes place. The $+$ and $-$ signs indicate the sign of the wavevector.}
\label{scat}
\end{figure}

Following our late time outgoing wavepacket backwards in time, scattering will occur both on the way in towards the hole, and on the way back out, 
contributing to the presence of components in the early time ingoing 
wavepacket with wavevectors $k_{-},\, k_{-s}, k_{+}$ (\ref{roots}). The different 
contributions to the scattered wave can be visualized easily if we
idealize the scattering as a process that occurs at just one position,
as depicted in Figure \ref{scat}.
(In fact this will be essentially the case for the ``kinked" metric 
to be discussed in section \ref{Results}.)
Going backwards in time, the $k_{+s}$-packet back-scatters into the other three 
roots. A $k_{+s}$ piece continues on towards the horizon,
blueshifting, and undergoes mode conversion to a pair of $k_{-}$ and $k_{+}$
packets, which then propagate back away from the horizon. When these reach
the scattering ``kink", they partly back-scatter into the $k_{+s}$
root and forward scatter into the other three roots. 
The $k_{-}$ packet, which determines the particle creation amplitude,
thus receives contributions from three independent processes: 
(i) backscattering from $k_{+s}$, (ii) mode conversion at the horizon
(partly modified by forward scattering out of $k_{-}$), 
and (iii) forward scattering from $k_{+}$.

For the ordinary wave equation in 
more than two dimensions, there is also backscattering from the geometry.
However in that case there is only one ingoing mode for fixed positive
$\omega$, corresponding to $k_{-s}$, which has positive free-fall
frequency and does not give rise to any particle creation. In our
model the possibility of scattering into the $k_{-}$
mode, which has {\it negative} free-fall frequency (and hence negative
norm), does lead to particle production! This particle creation has
absolutely nothing to do with black holes. It is caused by scattering from a stationary (in fact static) geometry. This is possible because the time translation Killing field
does not agree with the free-fall frame which is a distinguished 
frame in our Lorentz non-invariant theory. We shall have more to say about 
particle creation in a stationary metric in section \ref{puzzle}. 

\section{Computing the spectrum of black hole radiation}
\label{method}

 In this section we discuss the approach taken to solve for the particle
production in a given mode, i.e., fixed Killing frequency solution.  We
begin by discussing the reduction of the PDE, equation (\ref{eom}), to an ODE
plus boundary condition.  We then discuss the method applied to compute 
the particle prodution rate given a solution to the ODE.

\subsection{Solving the mode equation}
\label{sec:ode}
Because the background flow is stationary, the Killing frequency $\omega$
is conserved, and solutions to equation (\ref{eom}) of the form 
$\phi(t,x) =e^{-i \omega t} f(x)$ exist.  Substitution yields an ODE for
$f(x)$, 
\begin{equation}
(-i \omega + \partial_{x} v)(-i \omega + v\partial_{x})f(x) =
\hat{F}^{2}(\partial_{x})f(x).
\label{ode}
\end{equation}  
Boundary conditions are required to select among 
solutions of this equation. We would like to impose boundary conditions
that hold if the solution arises as part of a wavepacket that is localized and outgoing at late times, since these are the ones that are relevant for the
Hawking effect.  What are these boundary conditions?

Note first that a solution cannot strictly vanish everywhere inside
the horizon unless it is identically zero everywhere or it is somehow singular at the horizon. Since our PDE contains higher spatial derivative terms, and is hence not causal, one might expect that solutions would be well behaved at the horizon. This is in fact the case, and therefore a solution cannot vanish everywhere inside the horizon. Unruh's results from propagating wavepackets
satisfying a similar equation indicate that outgoing wavepackets do indeed 
have non-vanishing amplitude inside the horizon, but they decay rapidly 
inside.\footnote{We thank W.G. Unruh for showing us some of his unpublished graphs that make this point clear.} We are thus interested in the mode 
solutions to the ODE that decay inside the horizon.

For example, consider the case where the free-fall velocity $v(x)$
goes to a constant (less than $-1$) inside the horizon. In this constant
velocity region the ODE (\ref{ode}) has 
two oscillating modes, one exponentially growing mode, and
one exponentially damped mode.  In principle we apply the boundary condition that the
solution and its first three derivatives
agree with the exponentially growing (with $x$) mode and its first three derivatives respectively at some position in the constant velocity region
inside the horizon.  As a practical matter however, it is very important to 
realize that almost any boundary condition applied {\it inside} the horizon will
yield essentially the same solution {\it outside} the horizon. This is
important because for a general $v(x)$ we would not know explicitly the
form of the required boundary condition, and also because numerical
integration of the ODE will always introduce small errors.

The insensitivity to the boundary condition can be understood as follows.
An arbitrary boundary condition may be written
as a linear combination of the boundary conditions generating the above
four modes.  Since the solution corresponding to the growing mode grows
exponentially with increasing $x$ until it emerges from behind the horizon,
whereas the other three maintain approximately constant or decreasing
amplitudes, the contribution from the exponentially growing mode will
be exponentially amplified relative to the other modes. 
Moreover, agreement (up to an irrelevant overall constant) outside the horizon with the solution arising purely from the exponentially damped boundary condition will rapidly get better and better as the arbitrary condition is applied deeper and deeper inside the horizon. In fact, in numerically
solving the ODE, we used the criterion of insensitivity to the point of application of the boundary condition to ensure that the appropriate mode
had been selected. Also, it was important not to start the numerical 
integration too deep inside the horizon, for otherwise the exponential
growth would produce such huge field values that numerical problems arose.

\begin{figure}[h]
\centerline{
\psfig{figure=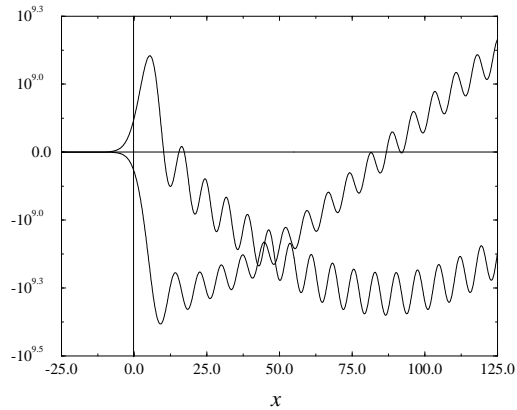,angle=-90,height=6cm}}
\caption{\small Plot of the real and imaginary parts of the solution to 
eqn. (\ref{ode}) for a free-fall velocity $v_{\rm kink}$ (\ref{vkink}), temperature $T_{H}$ = 0.003, and a frequency of $\omega$ = 0.01, in units where $k_0=1$.  
The horizon is located at $x=0$ and the kink is located at $x\simeq 26$.  Note how the solution tunnels out across the horizon, growing exponentially, and then begins oscillating. Both the short and long wavelength components are
clearly visible.}
\label{soln}
\end{figure}

Given the above ODE and boundary conditions it is now a simple matter to
solve the equation numerically\footnote{For the numerical work we used
Mathematica for its convenience and flexibility. In retrospect, it would
probably have been better to use a standard programming language since the 
computations turned out to be very time consuming. For instance, computing
the created particle flux at the highest resolution for a {\it single} 
frequency took on the order of one hour on a Next machine.} or, in some cases, analytically. 
(The analytical solution is given in the Appendix.)
We need only integrate the equation out to positive $x$ values in which
the velocity is approximately constant (and at its asymptotic value).  In
a region where $dv/dx$ is zero (or is negligibly small) the generic mode solution 
$u_\omega(x)$ is just a sum of four harmonics,
\begin{equation}
u_\omega(x)\rightarrow\sum_{l=1}^{4} c_{l}(\omega) e^{ik_l(\omega)x},
\label{harmonics}
\end{equation}
with $k_l(\omega)$ given by (\ref{roots}). An example solution is shown in Figure \ref{soln}.  As expected, the solution grows exponentially out across the horizon, located at $x=0$, and then starts oscillating.  
By fitting (i.e., employing a minimization procedure) for the
coefficients  $c_{l}(\omega)$ of the four mode solutions, the numerical
solution can be expressed analytically in the asymptotic constant velocity region.
This step amounts to taking a ``local Fourier transform" of the numerical
solution in the constant velocity region.  As will be shown below, 
knowledge of these coefficients is tantamount to finding the particle 
production in mode $\omega$.
As a check on the accuracy of the numerical solution we also used the exact solution (\ref{exsoln}) (given in the Appendix) to compute the particle production at a given $\omega$,
and found agreement to within the targeted numerical accuracy.

\subsection{Computing the particle creation rate}
\label{pc}
In this subsection we obtain the explicit expression for the particle creation rate to be used in our analysis. Let
$\psi_{{\rm out}}$ denote a final wavepacket of the form 
\begin{equation}
\psi_{\rm out}={\cal N} 
\int d\omega\, c_{+s}(\omega)\, 
 e^{i \omega u_{o}}\, e^{-i (\omega t - k_{+s}(\omega) x)},
\label{outpacket}
\end{equation}
where ${\cal N}$ is a normalization constant, 
$u_{o}$ is a constant that determines at what time the wavepacket
passes a given point, and only the small positive root $k_{+s}(\omega)$
of the dispersion relation
is included in the integration.  This is an outgoing wavepacket solution
of the field equation (\ref{eom}) in the constant velocity region,
where it has both positive Killing frequency and positive free-fall frequency.
 
Propagating this data backwards in time, as explained in section 
\ref{wpandmc}, it arrives back again in the constant velocity region as an ingoing wavepacket
\begin{equation} 
\psi_{\rm in}=\psi_{-}+\psi_{-s}+\psi_{+}
\label{psiin} 
\end{equation}
composed of wavevector components $k_{-},\, k_{-s}\, k_{+}$
respectively, all of which have negative group velocity. 
Since the inner product (\ref{inner}) is time-independent we have
\begin{equation}
(\psi_{{\rm out}},\hat{\phi}) =  (\psi_{-},\hat{\phi})+ 
(\psi_{-s},\hat{\phi}) + (\psi_{+},\hat{\phi}) 
\end{equation}
or, in terms of annihilation and creation operators 
(\ref{C2}) and (\ref{C2*}),
\begin{equation}
a(\psi_{{\rm out}}) = - a^{\dagger}(\psi_{-}^*)+ a(\psi_{-s})+ a(\psi_{+}) .
\end{equation}
We assume that the state of the field at early times is the free-fall vacuum, $|{\rm ff}\rangle$, which satisfies
$a(p_{\rm in})|{\rm ff}\rangle$ = 0 for any ingoing positive free-fall 
frequency wavepacket $p_{\rm in}$. The particle creation in the packet 
$\psi_{{\rm out}}$, characterized by the expectation value of the number
operator, is thus (using (\ref{C3})) given by the norm of $\psi_{-}$:
\begin{equation}
N(\psi_{\rm out})= \langle{\rm ff}|
a^{\dagger}(\psi_{{\rm out}})a(\psi_{{\rm out}})
|{\rm ff}\rangle = 
-(\psi_{-},\psi_{-}).
\label{number}
\end{equation} 
To evaluate this norm we use the solution to the mode equation as follows.
 
Let $\psi$ denote the solution to equation (\ref{eom}) generated
by the final data set $\psi_{{\rm out}}$ (and therefore also by the early 
data set $\psi_{\rm in}$). This solution can be expanded in exact
solutions to the mode equation (\ref{ode}), 
\begin{equation}
\psi(t,x) = \int d\omega\,  e^{i \omega u_{o}}\, e^{-i\omega t}\, u_\omega(x),
\label{solution}
\end{equation}
where $u_\omega(x)$ is proportional to the mode solution that decays inside the horizon which we find either analytically or numerically.
At large positive $x$, in the constant velocity region, $u_\omega$ takes the form 
\begin{equation}
u_\omega(x)=\sum_{l=1}^{4} c_{l}(\omega) e^{ik_l(\omega)x},
\label{modes}
\end{equation}
as explained in the previous subsection. The $k_{+s}$ 
term in (\ref{modes}) gives rise to the outgoing wavepacket 
$\psi_{\rm out}$ at late
times, and the $k_{-}$, $k_{-s}$ and $k_{+}$ terms give rise to 
the ingoing wavepackets $\psi_{-}$, $\psi_{-s}$ and $\psi_{+}$ 
respectively at early times. Each of these
wavepackets is separately localized in the large $x$, constant velocity
region. In particular, $\psi_-$ takes there the simple form 
\begin{equation}
\psi_{-} = \int
d\omega \, c_{-}(\omega) e^{i \omega u_{o}} 
e^{-i(\omega t - k_{-}(\omega) x)}.
\label{psiminus}
\end{equation}
Now because $\psi_-$ is localized in the constant velocity region,
its norm can be evaluated as if the constant velocity region extended
throughout all of space. On a constant $v$ background one has 
\begin{equation}
(e^{-i(\omega t - k_{l}(\omega) x)},
e^{-i(\omega t - k_{j}(\omega) x)})_{{\rm constant}\; v} 
= 4\pi\omega'(k_l(\omega)) \delta (k_{l}(\omega) - k_{j}(\omega)),
\label{norm}
\end{equation}
where $\omega'$ is the free-fall frequency (\ref{otrans}). Using (\ref{norm}),
the number expectation (\ref{number}) evaluates to 
\begin{equation}
N(\psi_{\rm out}) = 4\pi\int 
d\omega \, |\omega'\bigl(k_-(\omega)\bigr)\,  
v_{g}\bigl(k_-(\omega)\bigr)\,  c_{-}(\omega)^{2}|.
\label{N}
\end{equation}  

In practice, the mode solutions $u_\omega$ we work with are not normalized.
To evaluate the norm of $\psi_{\rm out}$ note that 
at late times $\psi_{\rm out}$ is of the same form as (\ref{psiminus}),
with the small positive root $k_{+s}(\omega)$ in place of $k_-(\omega)$.
Thus $(\psi_{\rm out},\psi_{\rm out})=(\psi_{+s},\psi_{+s})=
4\pi \int 
d\omega\,  \omega'\bigl(k_{+s}(\omega)\bigr)\,  
|v_{g}\bigl(k_{+s}(\omega)\bigr)\, c_{+s}(\omega)^{2}|$.
Dividing (\ref{N}) by this norm then yields the properly normalized
occupation number. For a wavepacket that is strongly peaked about 
frequency $\omega$ this yields the number expectation value
\begin{equation}
N(\omega)=\frac{|\omega'(k_-)v_g(k_-)c_-^2(\omega)|}
{|\omega'(k_{+s})v_g(k_{+s})c_{+s}^2(\omega)|}
\label{No}
\end{equation}

Equation (\ref{N}) gives the occupation number of a particular wavepacket. 
To obtain the number of particles reaching ${\cal I}^{+}$ per unit time
consider the complete\footnote{Let $f(x)$ be any function with 
Fourier transform $\tilde{f}(k)$. On the interval $(j\epsilon,j\epsilon+\epsilon)$, $\tilde{f}(k)$
can be expressed as a Fourier series $\Sigma_n a_{jn} e^{i2\pi nk/\epsilon}$.
Thus one has $f(x)=\int dk \tilde{f}(k) e^{ikx}
=\Sigma_j\int_{j\epsilon}^{j\epsilon+\epsilon} dk\, 
\tilde{f}(k) e^{ikx}
= \Sigma_j\int_{j\epsilon}^{j\epsilon+\epsilon}dk\, 
\Sigma_n a_{jn} e^{i2\pi nk/\epsilon} e^{ikx}
=\Sigma_{jn} a_{jn} p_{jn}(x)$, where
$p_{jn}(x)=\int_{j\epsilon}^{(j+1)\epsilon}dk\, 
 e^{i2\pi nk/\epsilon} e^{ikx}$.}
orthonormal set of late time outgoing 
wavepackets of the form 
\begin{equation}
p_{jn}={\cal N}\int_{j\epsilon}^{(j+1)\epsilon} dk\, 
e^{-i\omega(k)t}\, e^{ikx}\, e^{i2\pi nk/\epsilon}.
\end{equation}
These are of the same 
form as the wavepackets (\ref{outpacket}), the only difference being in the 
integration over $k$ instead of $\omega$.
(This is the same basis as that used by Hawking\cite{Hawking},
except that for us $\omega(k)$ is a different function.) 
The wavepacket 
$p_{jn}$ is localized on the trajectory satisfying $v_g t - x - 2\pi n/\epsilon=0$, and the time interval between the passage of succesive
wavepackets $p_{jn}$, $p_{j,n+1}$ is therefore $\Delta t=2\pi/v_g\epsilon$.
Furthermore, the temporal spread of a packet at fixed $x$ is of the 
same order. The frequency spread $\Delta \omega$ in one of these wavepackets
is given for small $\epsilon$ by 
$\Delta \omega \approx (d\omega/dk) \,\epsilon = v_{g} \epsilon$. Thus
in the small $\epsilon$ limit we have $\Delta t\Delta \omega = 2\pi$.

The number of particles reaching ${\cal I}^{+}$ in a time interval 
$\delta t \ll \Delta t$ in the wavepackets $p_{jn}$ for fixed $j$ 
is approximately
\begin{equation}
\delta N(\omega_{j}) = \frac{\delta t}{\Delta t} N(\omega_{j}),
\end{equation}
where the notation $\omega_{j} \equiv \omega (j\epsilon)$ is introduced.
The approximation made is to assume that $\epsilon$ is very small so that
essentially only one packet for a given frequency contributes to the flux, 
and of that packet only the portion located in the time interval $\delta t$.  Rewriting this expression using $\Delta t\Delta\omega=2\pi$ and taking the
limit $\epsilon\rightarrow 0$ we obtain the particle number flux per unit frequency interval
\begin{equation}
\frac{dN(\omega)}{dtd\omega} = \frac{1}{2\pi} N(\omega) 
\end{equation}  
where we have dropped the subscripts.
In the thermal case, one has $N(\omega)=(e^{\omega/T}-1)^{-1}$.
The total luminosity, i.e. the total Killing energy reaching 
${\cal I}^{+}$ per unit time, 
is obtained from $dN(\omega)/dt$ by multiplying by the frequency, $\omega$, and summing up over all frequencies, i.e.,
\begin{equation}
L =  \int d\omega\,  \frac{\omega}{2\pi}\,  N(\omega).
\label{L}
\end{equation}

\section{Results of the computations}
\label{Results}

In this section we describe and interpret the results of the numerical computations we have performed.
Before discussing the results we introduce the relevant parameters.
Our models are completely characterized by the cutoff wavevector $k_0$
appearing in the dispersion relation (\ref{disp},\ref{F}) 
and the the free-fall velocity function $v(x)$ appearing in the
metric (\ref{metric}). (For a Schwarzschild black hole in 
4-dimensions $v(x)$ is given by $v(x)=-(2M/x)^{1/2}$.)
The most important parameter characterizing $v$ is the surface gravity
$\kappa=v'(x_h)$ where $x_h$ is the position of the horizon 
($v(x_h)=-1$), in terms of which 
the Hawking temperature is $T_H=\kappa/2\pi$. For the results discussed 
here, the ratio $T_H/k_0$ was either 0.000784 or 0.00312 (which for 
brevity we usually write as 0.0008 or 0.003 respectively), except
in one case where we considered a large number of temperatures
(see Fig. \ref{detcoeff}).
Note that if $k_0$ is the Planck mass then these temperatures are extremely
high and would correspond to very small Schwarzschild black 
holes.\footnote{It would have been interesting to compute the spectrum 
also for large
black holes, for example for a primordial black hole temperature
of, say,  $T_{H} \approx 10^{-20} k_0$. Such low temperatures 
lead to computational difficulties with the techniques we have employed
for two reasons. First, the distance from the horizon to the constant
velocity region extends over a width determined by $1/T_H$ which in units
of the length $k_0^{-1}$ is $k_0/T_H\sim 10^{20}$. It would therefore be 
necessary to integrate the ODE over an extremely large number
of oscillation cycles to obtain the particle production rate for a
typical frequency. This difficulty might be sidestepped by taking
the asymptotic velocity to be $-1+\epsilon$ where $\epsilon$ is of order $10^{-20}$. The second difficulty is simply that one would
have to design a code specifically to handle a computation in which 
numbers with ratios like $10^{20}$ are both important.}

Besides the surface gravity, what is important about $v(x)$ is how much
scattering (reflection) it produces. This is determined by the specific
form of $v(x)$. The results discussed here were obtained with 
three different forms for $v(x)$:
\begin{equation}
 v_{{\rm tanh},n}
\equiv {\textstyle{1\over2}}(\tanh((2\kappa x)^{n}))^{1/n} - 1
\end{equation}
for $n$ = 2 and 20 and
\begin{equation}
v_{\rm kink}(x)\equiv (\kappa x \theta(-x) - {\textstyle{1\over2}}),
\label{vkink}
\end{equation}
all of which go to $-1/2$ as $x$ goes to infinity.  
Qualitatively speaking, $v_{\rm tanh,2}$ produces very little scattering 
whereas $v_{\rm tanh,20}$ and $v_{\rm kink}$ produce much more (although
still very little). The curvature scalar of the line element (\ref{metric})
is given by $R=2(v^2)''$, so as $n$ increases in $v_{{\rm tanh},n}$ 
the curvature at the bend diverges, as it does for $v_{\rm kink}$.
{\it Important note:}
We shall often refer to the  $v_{\rm tanh,20}$ case as ``kinked", together with the truly kinked case $v_{\rm kink}$, since they produce comparable amounts of scattering. We considered them both because (i) the truly kinked case 
allowed us to check the numerical computation by comparing with the
analytic solution discussed in the Appendix, (ii) comparison of the
two allowed us to rule out spurious effects that might have been associated with a true kink,
and (iii) the comparison gave us more information about the impact of 
short wavelength processes on the spectrum of created particles. 
The most dramatic difference we found is illustrated by the two 
plots on the left hand side of Figure \ref{fourplots}.

The highest frequency at which there exist outgoing modes at the
asymptotic velocity $v=-1/2$ is given by $\omega_c \approx 0.18 k_0$. 
(At this value of $\omega$ the larger negative $k$ root of the dispersion relation still corresponds to a negative free-fall frequency (see Fig. \ref{dispreln}).) The ratio $\omega_c/T_H$ for the two temperatures used in our computations is about 235 and 59 for $T_{H}/k_{0}$ = 0.000784 and 0.00312 respectively.

With the introduction of the new length scale $k_0^{-1}$
one expects necessarily some deviation from the thermal
Hawking prediction. We shall use
$T_H/k_0$ as a benchmark with which to compare deviations. 
The initial deviation is presumably of order $(T_H/k_0)^p$ 
for some $p$. 
As stated in the introduction, previous analyses estimated
that $p$ should be unity or possibly smaller. 
(For the smooth metric it turns out that $p$ is 3 at the lower temperature,
and is roughly unity for the kinked metrics.)

The computations were performed using Mathematica. In order to test
reliability of the computations, we adjusted the WorkingPrecision,
AccuracyGoal, and PrecisionGoal options until there was no significant
change in the results. In all cases the WorkingPrecision
was better than $10^{-4}$ times the ratio of the smallest to the largest
coefficient in the Fourier expansion (\ref{harmonics}). We therefore
believe the results reported below accurately represent the solution to 
the model.
 
\subsection{Results}

In broad terms, the results are the following. 
For the smooth metric there is spectacular agreement 
with the thermal Hawking flux at low frequencies.
The relative difference from the thermal prediction\footnote{By the 
``thermal prediction" we mean $N(\omega)=(e^{\omega/T}-1)^{-1}$.
Any modifications due to ``grey body factors" are neglected.}
does not exceed $T_H/k_0$ until $\omega/T_H$ is of order 
$10^2$. The largest deviations reliably obtained were
of order 60\% at the cutoff frequency in the high temperature 
case. The kinked metric cases behaved quite differently. 
For them there is a significant non-thermal contribution
to the particle creation from scattering at the kink.
This contribution does not decrease with frequency so 
at some frequency it begins to dominate and the
relative deviation from the exponentially dying thermal
prediction grows exponentially. Furthermore, it oscillates 
as a function of frequency due to interference
between different scattered pieces. 
These results will now be described in detail.
We first compare the calculated flux of particles as a function 
of frequency to the thermal prediction.  
Next, we investigate cumulative effects of this deviation in 
terms of the total luminosity.  

\subsubsection{Flux - Smooth}
\label{smooth}

First we discuss the results for the smooth metric. 
Figure \ref{Table2} shows the relative deviation of the 
computed flux from the thermal flux for various 
frequencies up to a frequency around which the 
relative deviation exceeds $T_H/k_0$. Note the 
spectacular agreement between the computed and thermal
fluxes. For the lower temperature the relative deviation
at $\omega\simeq T_H$ is of order $(T_H/k_0)^3$, while for 
the higher temperature it is of order $(T_H/k_0)^2$,
both significantly less than the expected deviation of
order $T_H/k_0$. Since we have only two fairly high
temperatures however it is difficult to say what the 
real dependence on $T_h/k_0$ is. The size of the deviation
could be small in part due to a small coefficient.
It should be possible to extract the temperature dependence by 
evaluating the relative deviation at a fixed value of $\omega/T$
for several values of $T$. This is difficult because of the small
numbers in the smooth case, but the analogous calculation has
been performed for the kinked case, as described in the next subsection. 

A crude and very conservative estimate of the numerical error in the 
computation suggests that the low temperature relative differences 
in Figure \ref{Table2}
are of the same order of magnitude as the computational error.
(For the smallest frequencies, the relative difference is just
very small, while at larger frequencies the flux itself is so 
small that it is difficult to evaluate the relative difference 
accurately.) 
This is probably an overestimate of the error, but if correct it means that 
only the order of magnitude of the low temperature numbers is significant.

For both temperatures we see an oscillation in the sign of the
relative difference as a function of frequency.

\begin{figure}
\centerline{
\begin{tabular}{|l|l||l|l|}                        \hline\hline
\multicolumn{2}{|c||}{$T_{H}/k_{0}=0.0008$} & 
  \multicolumn{2}{|c|}{$T_{H}/k_{0}=0.003$} \\ \hline
$\omega/k_{0}$ & Rel. Diff. & $\omega/k_{0}$ & Rel. Diff. \\ \hline\hline
0.00049 & -2.2 $10^{-9}$  & 0.0020  & -8.2 $10^{-7}$ \\ \hline
0.0054  & 1.9 $10^{-8}$   & 0.0098  & -1.0 $10^{-6}$ \\ \hline
0.010   & 2.0 $10^{-7}$   & 0.018   &  2.1 $10^{-6}$ \\ \hline
0.015   & 6.9 $10^{-7}$   & 0.025   &  1.2 $10^{-5}$ \\ \hline
0.020   & 1.5 $10^{-6}$   & 0.041   &  6.1 $10^{-5}$ \\ \hline
0.025   & 2.0 $10^{-6}$   & 0.057   &  1.3 $10^{-4}$ \\ \hline
0.030   & -1.5 $10^{-5}$  & 0.072   & -5.6 $10^{-6}$ \\ \hline
0.035   & -2.4 $10^{-4}$  & 0.088   & -1.3 $10^{-3}$ \\ \hline\hline
\end{tabular}}
\caption{\small Relative difference between computed and thermal energy fluxes for a range of low frequencies for both values of the Hawking temperature.}
\label{Table2}
\end{figure}

\begin{figure}[tb]
\centerline{\psfig{figure=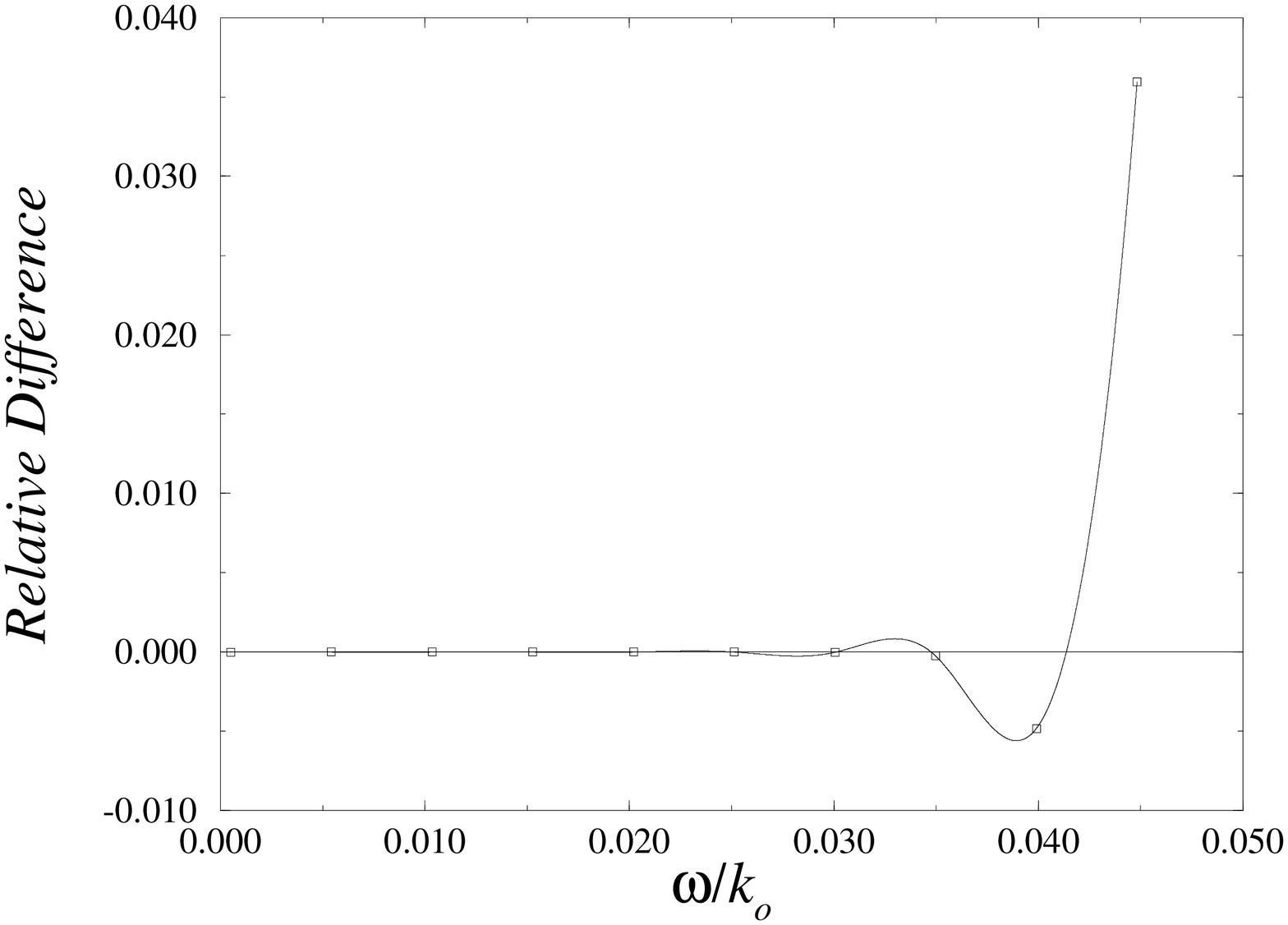,angle=-90,height=6.5cm}\hfill
\psfig{figure=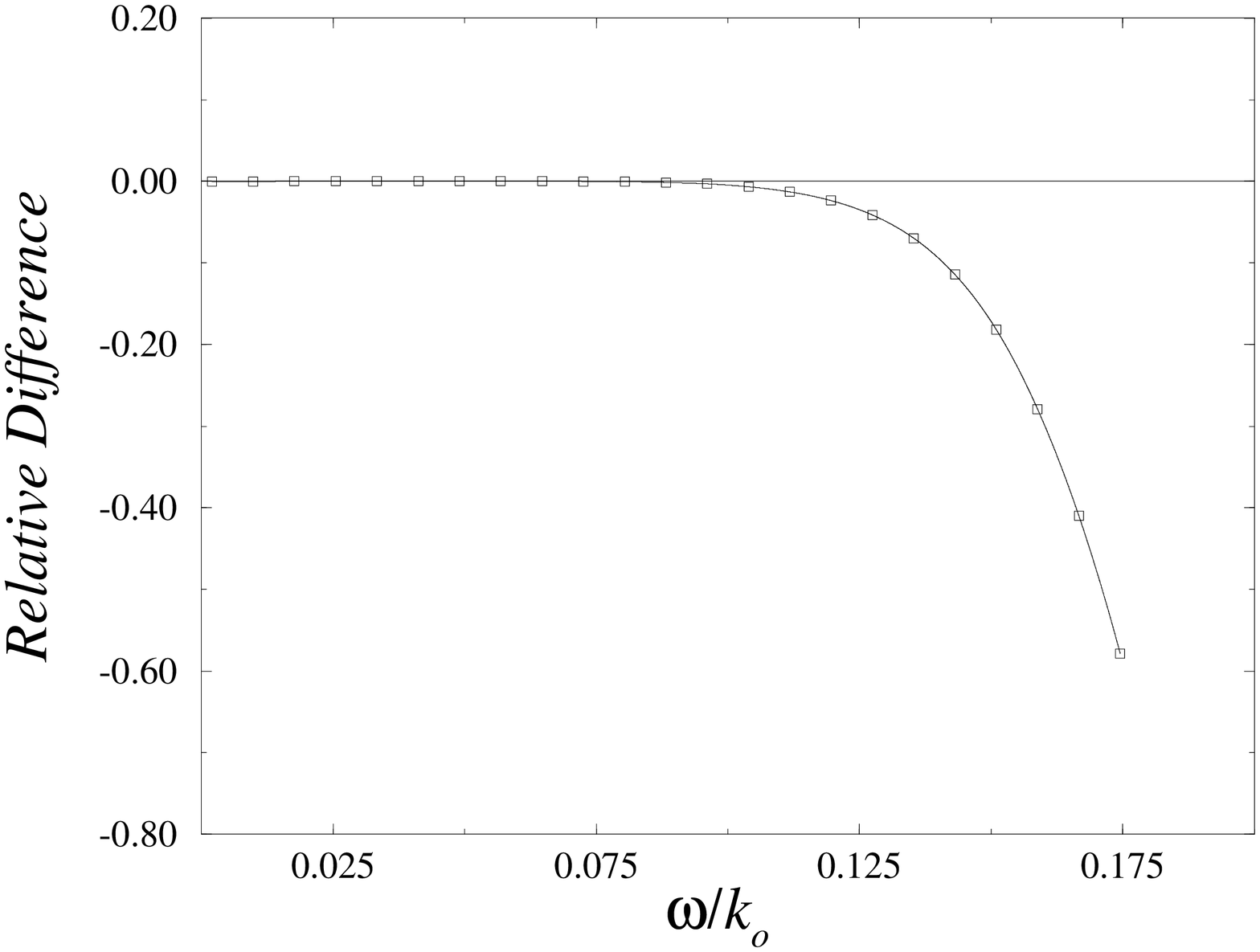,angle=-90,height=6.5cm}}
\caption{\small Plots of the relative difference between computed and thermal energy fluxes as a function of $\omega/k_{0}$ for the smooth metric.
$T_H/k_0$ is 0.0008 on the left and 0.003 on the right.}
\label{twoplots}
\end{figure}

The high frequency results are plotted in Figure \ref{twoplots} in the form of the relative difference between computed and thermal energy fluxes as a function of $\omega/k_{0}$ (the low frequency data is included but is too small to be observed from the graph).  For the higher temperature case the entire spectrum was obtained, and the maximum relative difference is about 60\% and occurs at 
the highest frequency. For the lower temperature case the agreement between the computed and thermal energy fluxes remained quite good out until $\omega/T_{H} = 57.2$ (the last data point obtained) where the relative error was about 3\%.  Data could not be obtained accurately beyond this value of $\omega$ because
the coefficient of the negative norm mode was too difficult to compute accurately due to its extremely small size.

\subsubsection{Flux - Kinked}

\begin{figure}[tb]
\centerline{
\psfig{figure=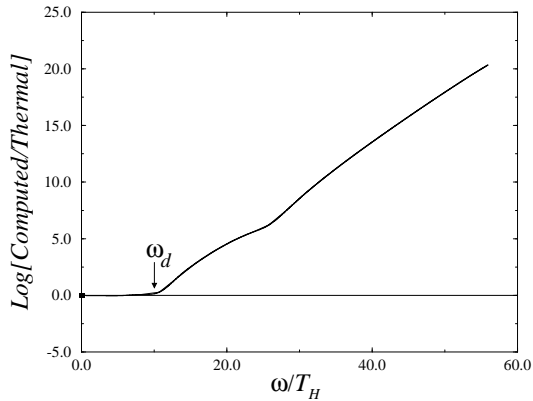,angle=-90,height=6cm}}
\caption{\small Plot of the logarithm of the ratio of computed flux reaching infinity to the thermal flux reaching infinity for $v_{\rm kink}$ and 
$T_{H}/k_{0}$ = 0.003, as a function of frequency in units of the Hawking 
temperature.  The position of the deviation frequency, $\omega_{d}$, for this case is indicated 
on the figure.}
\label{Logplot}
\end{figure}

Now we turn to the results for the kinked metrics.
In Fig. \ref{Logplot} 
we plot the logarithm of the ratio of the computed flux reaching infinity to the thermally predicted flux reaching infinity
as a function of frequency for a free-fall velocity $v_{\rm kink}$ and temperature $T_{H}/k_{0}$ = 0.003.  
This graph illustrates the generic form of deviations between the two
for all kinked cases. Namely, there was reasonable agreement 
(see below for a more precise statement) up to some frequency $\omega_{d}$, 
while above $\omega_{d}$ the disagreement was large and increased 
(on average, see below) with frequency. The frequency $\omega_{d}$, termed the deviation frequency, represents the approximate frequency where rapid growth (apparently exponential) of the ratio between the computed and thermal energy fluxes begins.
 
\begin{figure}[h]
\centerline{
\psfig{figure=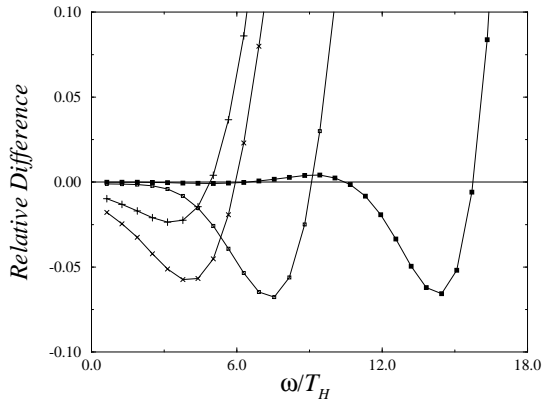,angle=-90,height=6cm}}
\caption{\small Plot of the fractional difference between computed flux 
and thermal flux as a function of frequency in units of the Hawking temperature. The curves correspond from left to right to the $(T,v)$ pairs 
(h,$v_{{\rm tanh},20}$), (h,$v_{\rm kink}$), (l,$v_{\rm kink}$), 
(l,$v_{{\rm tanh},20}$), where l and h denote temperatures 0.0008$k_0$ and 
0.003$k_0$ respectively.}
\label{lowfreq}
\end{figure}

The low frequency deviations are magnified in Figure \ref{lowfreq},
in which the relative difference between the computed and thermal
fluxes is plotted. 
At the lowest frequencies, the computed flux in the lower temperature cases has already approached very close to the thermal prediction, whereas in the 
higher temperature cases there is still a difference of a couple of 
percent at the lowest frequency computed. We have not computed for yet lower
frequencies because, as we integrated across the kink, the convergence
criterion used by Mathematica failed. (Since the thermal number flux $(e^{\omega/T}-1)^{-1}$ diverges as $\omega\rightarrow 0$, the relative
difference must vanish in that limit unless the absolute difference
is also diverging. If a scattered component survives as $\omega\rightarrow 0$,
then the absolute difference {\it will} diverge.)

Increasing $\omega$ from zero, we see that the computed 
flux generically starts out slightly below the thermal flux. Depending on the parameters, the relative deviation then either decreases monotonically, or oscillates about the thermal value, eventually reaching a maximum negative value after which it becomes
positive and never returns to zero.
The maximum negative deviation is about 7\% for the lower temperature
(0.0008$k_0$) cases and is 2.4\% or 6\% for the higher temperature
(0.003$k_0$) cases. 
For comparison, note that $T_{H}/k_{0}$ and $\sqrt{T_{H}/k_{0}}$ are
around 0.1\% and 3\% respectively for the lower temperature 
and are around 0.4\% and 6\% respectively for the higher temperature. 
Thus the maximum negative deviation appears to be of order 
$\sqrt{T_{H}/k_{0}}$ in these cases. 

\begin{figure}[h]
\centerline{
\psfig{figure=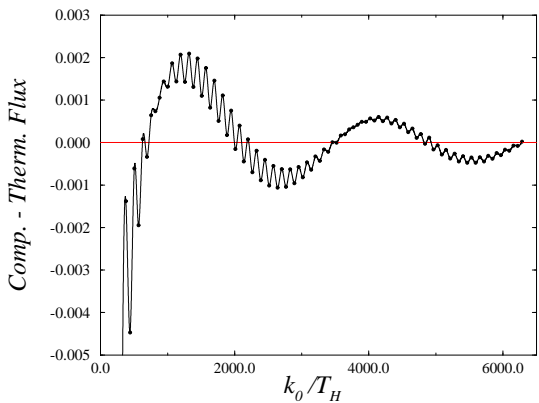,angle=-90,height=6cm}}
\caption{\small Plot of the difference of computed and thermal particle fluxes for $\omega=T_{H}$, as a function of $k_{0}/T_{H}$, for the kinked
metric $v_{\rm kink}$.}
\label{detcoeff}
\end{figure}

In Figure \ref{detcoeff} we investigate the convergence of the computed to the thermal particle number flux at the fixed frequency $\omega=T_H$,
for the metric $v_{\rm kink}$,
as the temperature is lowered (or, equivalently, as the cutoff is raised).
We see a small amplitude, short period oscillation superimposed on a
larger amplitude, longer period one. In order to characterize the rate of 
convergence we have found the absolute values at the four points where
the longer period oscillation peaks, and fit them to a power law
$(T_H/k_0)^p$. For all of the six pairs of points we find $p=1$
to within 10\%, so it appears that the difference is decaying like
$T_H/k_0$.

Past the last frequency at which the relative deviation crosses zero
the deviation takes off wildly. The computed flux ceases its exponential decrease at some $\omega_d$.  Fig. \ref{Logplot} shows that $\omega_d\sim 10 T_H$, whereas Fig. \ref{lowfreq} shows that the last zero occurs at around $6 T_H$
in that case.

\begin{figure}[t]
\centerline{
\begin{tabular}{|l|l|l|}                        \hline\hline
metric & $T_{H}/k_{0} = 0.0008$ & $T_{H}/k_{0}=0.003$ \\ \hline\hline
$v_{\rm tanh, 20}$  & 16.5  &  10 \\ \hline
$v_{\rm kink}$  & 13   & 9.5 \\ \hline\hline
\end{tabular}}
\caption{\small  Deviation frequencies $\omega_d$, in units of the Hawking temperature, for the various kinked free fall velocities and temperatures, from an eyeball fit of
\ref{Logplot} (and the analogous plots for the other kinked metrics which are not included here) to the data.} 
\label{Table1}
\end{figure}

Before discussing the results for large $\omega$, we look at how $\omega_{d}$ varies with the temperature $T_{H}/k_{0}$ and free-fall velocity (i.e., form
of the metric). The values of $\omega_{d}/T_H$ are listed in Figure \ref{Table1}. Note that $\omega_{d}/T_{H}$ increases with decreasing 
temperature, and in the lower temperature case is somewhat larger 
for the smoothed kink than for the true kink. 

Looking now at the deviations at frequencies above $\omega_{d}$ we see from Figure \ref{Logplot} that the Log curve is approximately linear.
That is, one has 
\begin{equation}
\frac{(dN(\omega)/dtd\omega)}{(e^{\omega/T} - 1)^{-1}} \sim
f(\omega) e^{\gamma (\omega - \omega_{d})},
\end{equation}
where $f(\omega)$ is some relatively slowly varying (i.e. not exponential)
function.
We have estimated the value of $\gamma$  from graphs like Figure \ref{Logplot} for both kinked metrics. We find that $\gamma \approx 1/T$
to within the roughly 5\% accuracy of our estimate.
The value $\gamma=1/T$ is what one would have 
if the spectral flux were {\it constant}, so we conclude that the exponential
suppression that would be present in the thermal flux is arrested after 
$\omega_d$ in our model with the kinked metrics. Putting $\gamma=1/T$ we thus have,  for $\omega>\omega_d$, 
\begin{equation}
\frac{dN}{dt\,d\omega} \approx 
f(\omega)e^{-\omega_{d}/T}.
\label{fluxapprox}
\end{equation}

\begin{figure}[p]
\centerline{\psfig{figure=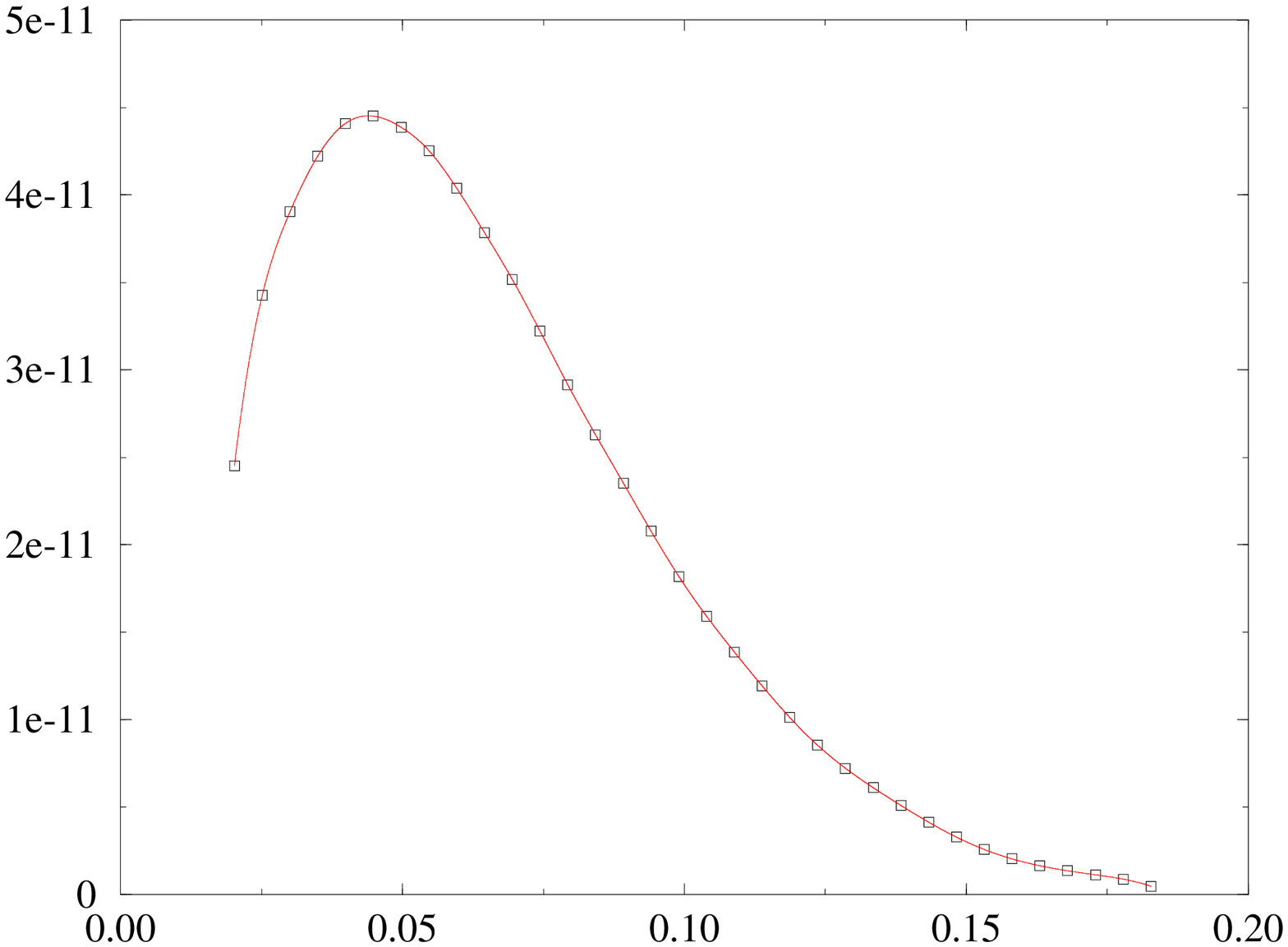,angle=-90,height=6.5cm}\hfill
\psfig{figure=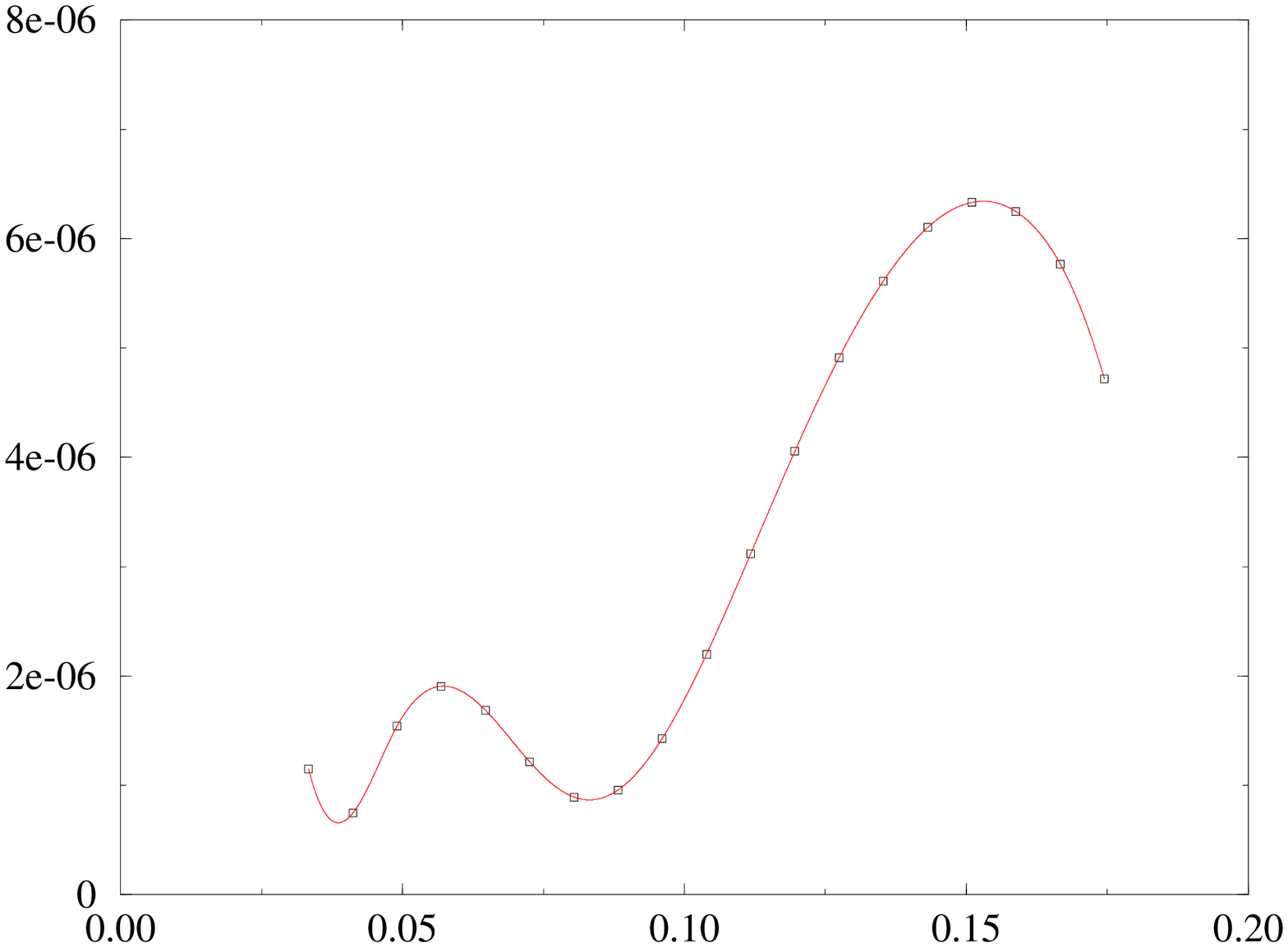,angle=-90,height=6.5cm}}
\centerline{\psfig{figure=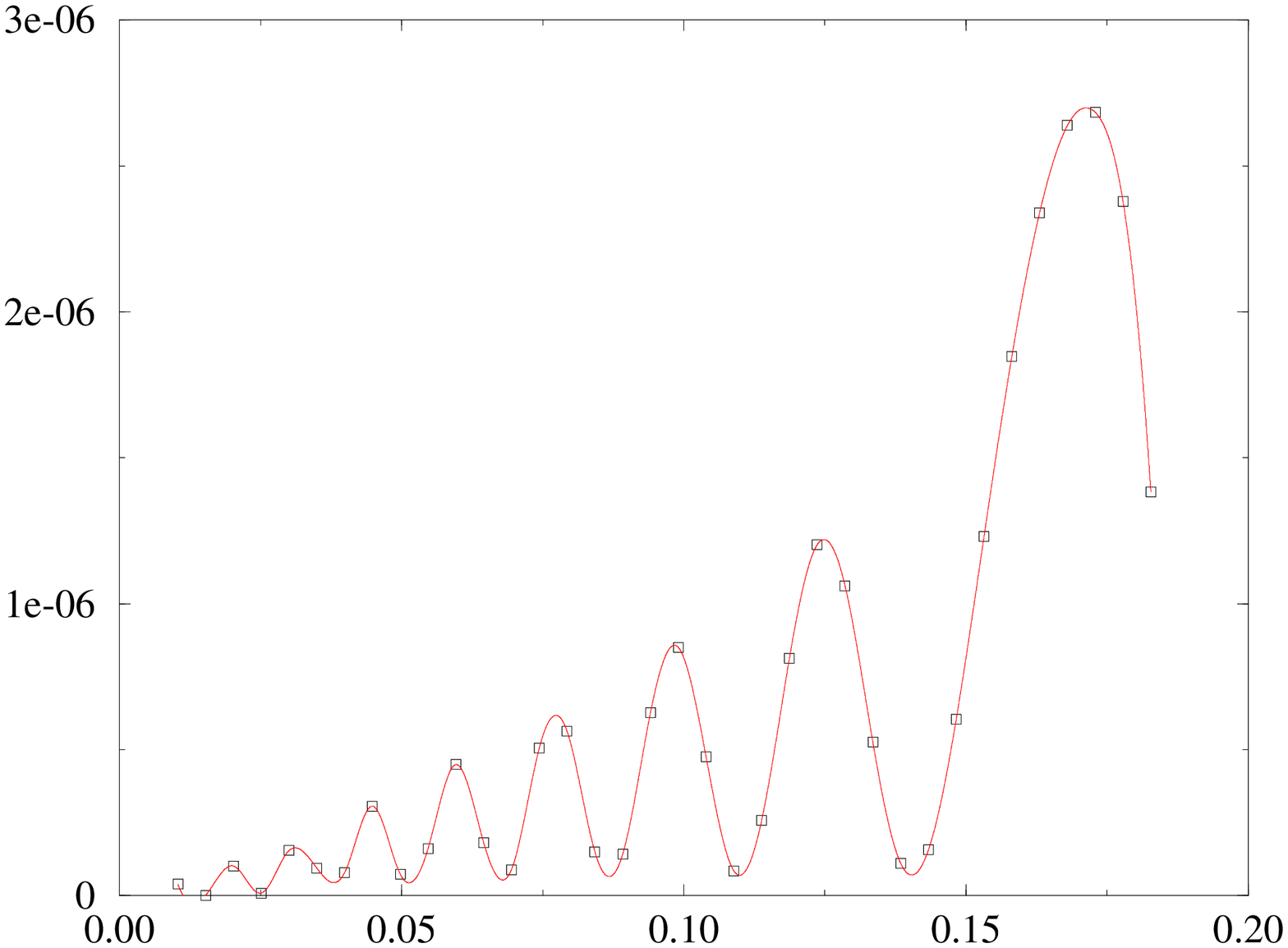,angle=-90,height=6.5cm}\hfill
\psfig{figure=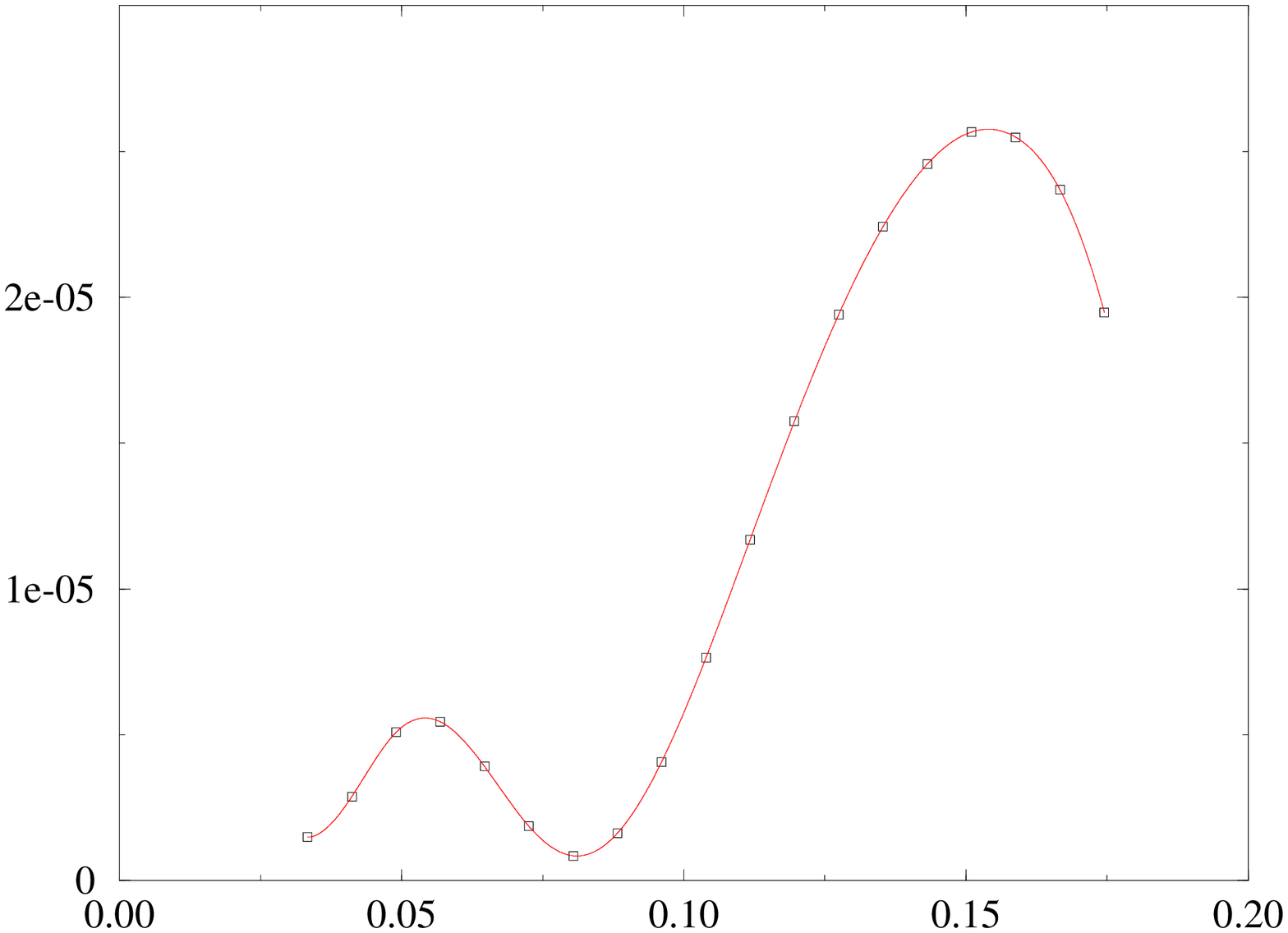,angle=-90,height=6.5cm}}
\caption{\small Plots of the energy flux as a function of frequency, both in units of the cutoff wavevector $k_{0}$, for all kinked cases.  The top row of plots is for the $v_{\rm tanh, 20}$ cases and the bottom row of plots is for the $v_{\rm kink}$ cases.  $T_H/k_0$ is 0.0008 on the left and 0.0003 
on the right.}
\label{fourplots}
\end{figure}

Actually there is interesting structure in the flux above $\omega_d$ that 
is ``washed out" in the logarithmic plot of Fig. \ref{Logplot}.
This structure is revealed in Fig. \ref{fourplots} which shows plots
of the computed flux of energy reaching infinity for frequencies larger than $\omega_{d}$ for the four kinked parameter sets. These plots show that in fact the 
energy flux does {\it not} increase linearly (as it would 
if the number flux were constant). In three of the four cases, the energy flux oscillates with frequency. The period of this oscillation increases with frequency, and the
maximum amplitude grows at least linearly and certainly faster in the $v_{\rm kink}$, $T_{H}/k_{0}=0.0008$ case. The minimum amplitude
is greater than zero and slowly growing.  Recall that even in the low frequency spectrum there was an oscillation of the computed flux about the thermal flux. Whether these two oscillations are related is an issue that will be
discussed in the interpretation subsection. In the other case, $v_{{\rm tanh},20}$
and the lower temperature $T_{H}/k_{0}=0.0008$, there is no oscillation
at all. Rather the flux increases, peaks,  and then decays again.
Nevertheless, it still remains many orders of magnitude above the thermal prediction even at the highest frequency.

\subsubsection{Luminosity}

Even a small deviation in the flux could have a dramatic effect on the
net luminosity if integrated over a wide enough range of frequencies.
What determines the range of frequencies in our case is that for 
$\omega$ greater than about 0.18$k_0$, there is no longer a solution to
the dispersion relation corresponding to the outgoing mode.\footnote{In Unruh's model\cite{Unruh2}, with the hyperbolic
dispersion relation, there is no cutoff frequency on outgoing
modes at all, so luminosity accumulates up to arbitrarily
high frequency. If the scattering phenomenon persists up to
arbitrarily high frequency then it appears there will be an
infinite energy density of created particles in that model. 
(Since the group velocity is dropping exponentially however this would not
translate into an infinite energy flux.) For a truly kinked
metric one would indeed expect the scattering to persist up
to arbitrarily high frequencies, but if the metric is slowly
varying below some length scale, then for wavelengths shorter 
than that scale one would expect the scattering to cease.}

For the smooth metric ($v_{{\rm tanh},2}$) with either temperature, as well as for $v_{{\rm tanh},20}$ with $T_{H}/k_{0}$ = 0.0008, we find no significant deviations from the thermal luminosity. 
For the other cases, however, there is a significant change.
To estimate the cumulative effects of the deviations in the energy flux we compute a lower bound to the portion of the 
luminosity (\ref{L}) in the modes of frequency larger than $\omega_{d}$.  
(Since the flux in our models is primarily {\it below} the thermal flux
at lower frequencies, including those would somewhat counterbalance the
extra flux at the higher frequencies.)
The simplest way to do this is to evaluate a Riemann sum of (\ref{L}) from $\omega_{d}$ to the largest allowed frequency.  In the thermal case this integral amounts to less than 0.2\% of the total luminosity.  
For $v_{{\rm tanh},20}$ and $v_{\rm kink}$ with $T_{H}/k_{0}$ = 0.003 we find
values of 2.6\% and 8\% respectively, and for $v_{\rm kink}$ at 
$T_{H}/k_{0} = 0.0008$ we find a value that is 10\% of the total thermal luminosity.

It would have been interesting to compute this extra luminosity for 
much lower temperatures (i.e. much larger black holes) but, as mentioned 
in a footnote at the beginning of this section, we are so far unable to handle very low temperatures. 
At low temperatures one might think that since the thermal flux is smaller,
the extra flux might make a larger relative contribution, since one is
always integrating up to the same cutoff frequency $\omega_c$ (since this is fixed for a fixed asymptotic value of $v(x)$).
But there are two other effects that are relevant. First, 
as the temperature falls, $\omega_d/k_0$ decreases (although $\omega_d/T_H$
increases, see Fig. \ref{Table1}), 
so the range from $\omega_d$ to $\omega_c$ increases,
which tends to increase the extra luminosity.
On the other hand, at lower temperatures, the kink is less kinky,
so produces fewer particles (which is reflected in the greater exponential suppression suggested by (\ref{fluxapprox})). 
We do not know the outcome of these competing effects.

\subsection{Interpretation of Results}

The goal of this section is to explain the various features of the results.  Listed below, and subdivided under the smooth and kinked headings, are the most important features that we shall address.  They are:
\vskip .2cm
\noindent
{\bf Smooth}

\begin{enumerate}

\item The extremely good agreement with thermality at small $\omega$.

\item The oscillations in the flux at small $\omega$; specifically, their origin and their characteristics (such as amplitude and period).

\item The energy flux behaviour for large $\omega$ for the higher temperature case.

\end{enumerate}

\noindent
{\bf Kinked}

\begin{enumerate}

\item The comparatively larger deviations from thermality at small $\omega$ in the energy flux than in the smooth metrics.

\item The form of the oscillating convergence to the thermal prediction
at $\omega=T_H$ as a function of $k_0/T_H$ (Fig. \ref{detcoeff}).

\item The dip in the relative difference between fluxes immediately before the very large deviations take over (Fig.\ref{lowfreq}).

\item The value of $\omega_{d}$.

\item The huge deviations from thermality at large $\omega$.

\item The oscillation in the energy flux at large $\omega$.

\item The lack of an oscillation in the lower temperature, $v_{\rm tanh, 20}$ metric at large $\omega$.

\end{enumerate}

The results for the smooth metrics are illustrated in Figures \ref{Table2} and \ref{twoplots}.   They show the extreme accuracy to which the computed particle creation agrees with the thermal prediction well out into the tail of the thermal spectrum.  Why is the agreement so good? {}From the discussion of wavepacket propagation in section \ref{wpandmc}, one might expect larger deviations than observed, since a wavepacket propagated backwards in time seemed to turn around and head back away from the black hole before ever reaching the horizon.  This might lead one to believe \cite{Ted1} that the appropriate temperature at which the black hole radiates a specific frequency mode should be $(1/2\pi)v'(x_t)$ where $x_t$ is the classical turning point. 
Since $v'$ is smaller at the turning point than at the horizon, one would 
thus expect {\it less} particle flux at a given frequency than in the thermal
spectrum at $T_H$. What we find however is that the computed particle production is {\it larger} than the thermal particle production for some low frequencies. 
Moreover, calculation reveals that the deviations shown in the table of Fig. \ref{Table2} are roughly
two orders of magnitude smaller than one would have expected if the appropriate
temperature were determined by $v'$ at the turning point. It thus appears that the packet must probe the horizon more closely than expected.  This is not so surprising since we have already seen that the WKB approximation breaks down around the classical turning point and hence so does our naive picture of the propagation of wavepackets in this region.  Furthermore, we know from our mode analysis that the packet does not vanish inside the horizon but rather decays exponentially across it. The small deviations that do exist may be due to the
scattering phenomenon discussed below, or may be attributable to an
intrinsic deviation from thermality of the mode conversion in these
models. We have not determined which of these is the case.

The remaining phenomena seem to be a consequence of scattering into 
and out of the negative wavevector, negative free-fall frequency mode,
principally from the kink in the metric. The scattering can be described using either a wavepacket picture, as described in section \ref{scattering},
or a time-independent mode picture. {}From the mode point of view, as the mode equation (\ref{ode}) is integrated out from the horizon,
a solution is obtained (Fig. \ref{soln}) with wavevector components
$k_{-}$, $k_{+s}$, and $k_{+}$ (\ref{roots}). 
(There is essentially no $k_{-s}$ component until the kink is reached,
due to the boundary condition inside the horizon and the fact that
mode conversion produces no such component.)
As the equation is integrated across the kink, the $k_-$ component 
receives contributions from (i) that already present on the left of
the kink (which is partly scattered), and (ii) the scattering of the 
$k_{+s}$ and $k_{+}$ components. The scattering is strongest from 
$v_{\rm kink}$, somewhat less strong for shorter wavelengths from 
$v_{\rm tanh,20}$, and barely present for the smooth metric 
$v_{\rm tanh,2}$. 

The contribution (i) is the
thermal contribution,  which arises from mode conversion at the horizon
as discussed in section \ref{conversion}. As the frequency $\omega$
grows, this contribution dies exponentially\footnote{We cannot rule out the possibility that there is a very small non-decaying piece. See section
\ref{smooth}.} as $\exp(-\omega/T_H)$. The contributions (ii) from the scattering will therefore dominate the particle creation for frequencies larger than some frequency, and they will not be exponentially decaying. This
is just what we found, as illustrated in Fig. \ref{Logplot}.
This interpretation can be checked by independently estimating the 
scattering amplitudes and predicting the deviation frequency $\omega_d$.

The scattering amplitudes for $\omega\gg T_H$ can be estimated as follows.
{}From the WKB approximation (i.e. using the local dispersion relation) 
we solve for either the $k_{+s}$ or the $k_{+}$ mode  
in the linear part of $v_{\rm kink}$ to the left of the kink. 
(If $\omega$ is not much greater than $T_H$ then the WKB approximation
is inadequate for the $k_+s$ mode.) In the constant part of $v_{\rm kink}$ to the right of the kink we know the exact harmonic modes, and the solution is a sum of these mods with constant coefficients.  The coefficients are determined by matching the solutions and their first three derivatives across the kink.  
{}From the coefficients one may then compute the amount of particle creation as was done in section \ref{pc}. 

Using this result we look for the value of $\omega$ where the particle creation from scattering is comparable to the particle creation from mode conversion.  For the higher temperature case ($T_{H}/k_{0}$=0.003) this yields $\omega_{d}/T_{H}$=9.5 and for the low temperature case ($T_{H}/k_{0}$=0.0008)  it yields $\omega_{d}/T_{H}$=13.  Looking back at the table in Figure \ref{Table1} we see extremely good agreement with the estimated values.  We cannot perform the same type of calculation with the $v_{\rm tanh,20}$ case, but of course in that case there is scattering as well, the amount of which would presumably be slightly less since now the discontinuity in $v'(x)$ has been smoothed out a little.  Hence $\omega_{d}/T_{H}$ would be slighty larger, 
in agreement with Figure \ref{Table1}.

The oscillation seen in the fluxes in the large $\omega$ spectrum for all kinked metrics except the low temperature $v_{\rm tanh, 20}$ case 
seems to be explained by interference between the contributions to the
$k_-$ mode from the scattered $k_{+s}$ and $k_{+}$ components. 
At these large frequencies the contribution of the mode conversion
is negligible, so essentially all the particle creation is coming from
the scattering.
Using the above WKB calculation we find that indeed the right order of magnitude is obtained to account for the peaks in the spectrum.  Furthermore, the amplitude of the $k_{-}$ pieces coming from the two separate scatters are the same order of magnitude. They will therefore produce significant interference
effects\footnote{If one takes a wavepacket with very small spatial extent at late times, this will produce narrow packets when propagated backwards in time.  Since the two scattering processes occur at different space-time points
(cf. section \ref{scattering}) the two scattered $k_{-}$ packets will
have essentially no overlap, so will not interfere. However, since we calculated the particle production with packets of perfectly well defined frequency 
(and hence spread out over all space), we will definitely have interference.}
as a function of frequency, since the relative phases of the two contributions
depend on frequency.

To check directly whether the interference explains the oscillation seen in the energy flux (see Figure \ref{fourplots}) we would need to know how the phases of the two $k_{-}$ contributions behave as functions of the Killing frequency of the late time outgoing packet.  We could easily extract the phase changes occuring throughout the propagation and at the scattering points from a WKB calculation, but we do not know how to calculate the phase change occuring during the mode conversion process. This could be sidestepped by calculating
instead the interference as a function of the position of the kink at 
a {\it fixed} frequency. Then the phase changes on mode conversion and
scattering would be essentially constant, so the oscillation would be 
due entirely to the difference in the phase accumulations of the 
$k_{+s}$ and $k_{+}$ modes as a function of position. We have not
perfomed the computations necessary to demonstrate that this is indeed
what is happening, but it seems quite plausible (and at this time the only plausible explanantion). The only potential problem we see  
is that the WKB predictions for the amplitudes of the two scattered 
modes, although of the correct order of magnitude, are not close enough 
to each other
to yield the large ratio of maximum to minimum flux values seen in Figure \ref{fourplots}. Since the WKB approximation gets better as $k$ grows,
it may be that the $k_{+s}$ contribution is not as accurately approximated
as the $k_{+}$ contribution. On the other hand, we currently see no reason
why the two contributions should in fact be as close in amplitude as they apparently are.

If the above interpretation is correct, it should also explain why
in the lower temperature $v_{\rm tanh, 20}$ case the energy flux 
at large $\omega$ is down by about five orders of magnitude from 
the other kinked cases, does not
oscillate but rather increases over some frequency range, and then decays
(see Figure \ref{fourplots}).  Recall that the bulk of the scattering occurs where $v'(x)$ changes from zero to $2 \pi T_{H}$.  In the $v_{\rm kink}$ case this occurs at a point, but in the  $v_{\rm tanh, 20}$ case it occurs over a range of approximately 20$k_0^{-1}$ for $T_H=0.00078$
and about $5k_0^{-1}$ for $T_H=0.0031$.  This smoothing 
seems to account for the difference in the spectra.  Since 
$k_{+}$ is of order $k_0$, the scattering for this mode is drastically reduced in the lower temperature case compared with the higher temperature case
because the wavelength is then much smaller than the length scale characterizing
the kink. Meanwhile, the amount of scattering of the $k_{+s}$ mode is
also greatly reduced, but for low enough Killing frequencies this reduction is not as severe since its wavelength is much longer. Hence, 
the scattering of the $k_{+s}$ mode dominates the particle creation
and there is no interference. 

Now let us consider the small $\omega$ oscillations. 
For small $\omega$, the thermal $k_{-}$  contribution from mode conversion is 
no longer negligible, so there are now three interfering contributions to the 
$k_{-}$ mode at large $x$. Since the amplitude of mode conversion 
dominates the other two we see only a small oscillation about this dominant contribution. At fixed $\omega=T_H$, the relative phases of the three
contributions are a function of the temperature. We expect that this 
explains the presence in Figure \ref{detcoeff} of the short period,
small amplitude oscillation superimposed on the longer period, larger 
amplitude oscillation, although we have not carried out any detailed
analysis to confirm this explanation. 

Our explanations of the remaining features are the most tentative.
The fact that the low frequency deviations are much greater in the 
kinked cases than the smooth case is presumably explained by the 
presence of scattering effects at these low frequencies. We could
not estimate these scattering amplitudes reliably using WKB since it is not valid for these low frequencies, and we have not attempted to compute
it in any other fashion. The oscillations in the flux at low frequency seen in the smooth cases and the lower temperature kinked cases are perhaps due 
to the same interference mechanism as we invoked to explain the 
high frequency oscillations. 
The dip in the relative difference of fluxes plotted in Figure \ref{lowfreq}
for the kinked cases is perhaps also an interference phenomenon arising from the scattering of wavepackets.  Around the frequencies at which this happens, the amplitudes of the scattered packets, as predicted from the WKB calculation, are of the same order of magnitude as the amplitude of the large negative wavevector packet arising from  mode conversion.  The question is why do the phases of these packets always conspire to produce this dip?  We unfortunately have no answer to this.

Finally, there is the large $\omega$ behaviour of the high temperature, smooth metric.  In contrast to all other cases, we get a computed energy flux 
{\it smaller} than the thermal energy flux, with a comparatively small relative difference of about 60\% (Figure \ref{twoplots}).  Perhaps this is related to the dip in the relative difference between fluxes discussed above. That is, perhaps in this case the dip has been pushed out to the end of the spectrum, so that the huge deviations that would have been present are now gone.  A possible test of this would be to see if one dip goes continuously over into the other as $n$ interpolates 
between 2 and 20 in $v_{\rm tanh, n}$, with the temperature fixed. 
This we have not done yet.

\section{The stationarity puzzle}
\label{puzzle}

The way the usual Hawking effect transpires in a strictly stationary spacetime
is that the outgoing wavepackets are traced backwards to parts that do 
{\it not}
make it back out to infinity, but rather cross the white hole horizon, 
at which
point the Unruh boundary condition on the quantum state is imposed. The 
piece of the wavepackets that scatters off the curvature and does make it back out to infinity is not associated with particle creation. Since in the Unruh
model and its variant considered here
the {\it entire} wavepacket turns around and goes back out,
it would seem that there can be no particle creation at all. 
So how do these models yield a nonzero Hawking flux?

The first answer is that neither we nor Unruh have actually followed the wavepacket
all the way out to infinity. Another way to put it is that we 
have taken the free-fall frame, in which the boundary condition is imposed, to 
be moving towards the black hole at infinity, rather than coinciding with the rest frame of the black hole at infinity. (Technically, this corresponds to
the fact that the metric function $v(x)$ goes to $-1/2$ rather than 
vanishing at infinity.)
But what happens if we take $v(x)$ to vanish at infinity, and continue to 
follow the wavepackets backwards in time, not stopping to impose the quantum state boundary condition until the packets reach infinity---or do something else? 
Do the wavepackets propagating backwards in time ever
reach infinity? Is there any Hawking radiation? 

This question can be addressed
as described in section 3
with WKB analysis, in which the motion of a wavepacket and the evolution of its wavevector is obtained by solving the local dispersion relation
$\bigl(\omega-v(x)k\bigr)^2=F^2(k)$ and integrating the group velocity.
The result of this sort of analysis depends on the 
particular dispersion relation considered. 
Here we will describe briefly some examples of the different types of 
behavior that seem to be possible.
As will become clear from the
discussion that follows, there are some ill-understood aspects of these
models, and we still lack a complete understanding of the physics 
in the presence of a true short distance cutoff.

\begin{itemize}
\item {\it Unruh model}:
In the Unruh model\cite{Unruh2}, with the hyperbolic tangent dispersion 
relation
$\omega=k_0\tanh(k/k_0)$, the magnitude of the wavevector grows without 
bound as the wavepacket moves outward where $v(x)$ is falling to zero.
Thus, even though the difference between the free-fall and Killing frames is
going to zero, the wavevector is diverging in such a way that the 
wavepacket always maintains a negative free-fall frequency part of the same,
negative, norm. Thus the Hawking effect indeed occurs as long as $v(x)$ never 
actually reaches zero. 
{}From this analysis we 
see that the Unruh model, while it entails a strict cutoff in free-fall frequency, involves in an essential way arbitrarily high wavevectors, i.e., 
arbitrarily short wavelengths.  Insofar as we wish to explore the consequences of a fundamental short distance cutoff on the Hawking effect, this is an unsatisfactory feature of the model. 
The outgoing modes emerging from the black hole region still arise from
arbitrarily short wavelength modes, albeit ingoing ones.

\item{\it Quartic model}:
This is the model we have focussed on in this paper, with dispersion relation
$\omega^2=k^2-k^4/k_0^2$ in free-fall coordinates. 
The spectrum of wavevectors associated with real frequencies is bounded at $k=\pm k_0$, and the group velocity goes through infinity and
changes sign at the cutoff wavevector. The
behavior of the positive and negative wavevector pieces (equivalently,
the positive and negative free-fall frequency pieces) of the early time
wavepacket are different. Proceeding backwards in time, the positive wavevector piece accelerates off to infinity at ever increasing velocity, exceeding the
long wavelength velocity of light. The negative wavevector part does the same
thing, except it appears from the WKB analysis that at some finite location, before the wavepacket reaches infinity, the wavevector reaches $-k_0$,
the free-fall frequency goes to zero, and the group velocity diverges.
The behavior at this point remains to be understood. There is a 
WKB solution that reverses direction in time, propagating ``back to
the future" and out to spatial infinity at superluminal velocity. 
However this solution violates the final boundary condition consisting
of the outgoing wavepacket with which we began the WKB analysis.
A solution that simply continues through with free-fall frequency changing
sign would have a norm that changes sign from negative to positive, which
is impossible in view of norm conservation.
It seems likely that the solution or its derivatives  must blow up
as the reversal point is apprached. This issue remains
to be clarified.

\item{\it Cubic Model}:
This model is defined by the dispersion relation $\omega^2=(k-k^3/k_0^2)^2$
in Lema\^{\i}tre coordinates. Note that in this model there is no bound to 
the size of $k$ for real frequency modes. The behavior of a backwards propagated low wavevector
outgoing wavepacket in this model is initially quite similar to that in the
original Unruh model and our quartic version thereof. As it recedes from the horizon (backwards in time) the positive norm 
part of the wavepacket propagates at increasing velocity
approaching close to twice the speed of light if $\omega\ll k_0$.
The negative norm part does the same for a while, but then as it approaches
zero free-fall frequency something new must happen, since norm conservation
does not permit the wavepacket to simply switch over to positive 
free-fall frequency. It seems likely that
the wavepacket undergoes partial mode conversion here to the negative 
frequency branch of the dispersion curve with $k<-k_0$. To keep the
norm conserved, this part would need to have a negative norm of larger
magnitude than before the conversion, since presumably a 
portion of the negative norm wavepacket continues on to positive
free-fall frequency and thus to positive norm. After this hypothetical conversion the negative norm piece has a group velocity of roughly twice the speed of light, and is heading (backwards in time) {\it toward} the horizon. 
It sails through the horizon at ever increasing velocity and slams into the singularity at infinite coordinate speed and with infinite wavevector.
To predict the outgoing Hawking spectrum would thus require a supplementary
boundary condition on these superluminal infinite wavevector modes at the singularity. If we take all this seriously and impose a vacuum boundary 
condition at the singularity, the predicted Hawking flux would depend on
how much of the negative norm wavepacket converted to positive norm at the
second conversion event.

\end{itemize}

It is clear from these examples that in facing the stationarity puzzle we 
are inevitably forced into the details of the short distance domain
of the model. For this reason it would be best to have a model
whose short distance behavior is determined by some reasonable physics.
One approach would be to push Unruh's original fluid analogy, taking 
for the fluid liquid helium, a fluid that can exist at zero temperature. 
This was considered initially in \cite{Ted1}, and the analysis has recently 
be pushed further in \cite{outgoing}. In this model 
the outgoing wavepacket is indeed traced backwards all the way back out
to infinity where it appears as a superposition of multi-roton 
modes of the superfluid. 

If indeed the outgoing modes arise from ingoing modes that start 
at infinity, then one is faced with the puzzle of explaining 
how particle creation could ever occur, in view of the conservation of 
the Killing energy. In \cite{outgoing} it is argued that the resolution
of this puzzle probably involves the back-reaction, which can both
destroy the Killing symmetry and decohere the positive and negative
norm parts of the ingoing wavepacket.

\section{Conclusion}
We have studied the spectrum of created particles in smooth
and ``kinked" two-dimensional black hole geometries for a linear, hermitian
scalar field satisfying a Lorentz non-invariant field equation
with higher spatial derivative terms that are suppressed by powers
of a fundamental momentum scale $k_0$. 
We have found that there are two qualitatively different
processes leading to particle production in this model. 
First, there is a thermal Hawking flux generated by 
``mode conversion" at the black hole horizon. The other process
has nothing to do with the horizon, and generates a non-thermal
spectrum via scattering off the background into negative free-fall
frequency modes. This second process does not occur for the ordinary
wave equation because such modes do not propagate outside the horizon 
with positive Killing frequency. 

The horizon component of the radiation is astonishingly close to 
a perfect thermal spectrum, as evidenced by our computations for
smooth metrics in which the scattered component is minimal. 
At $\omega/T_H=1$ the relative difference between the two is of 
order $(T_H/k_0)^3$ in the lower temperature case considered 
($T_H=0.0008k_0$), a much smaller difference than might have been expected.
Moreover, agreement to order $T_H/k_0$ persists out to 
$\omega/T_H\simeq 45$, where the thermal number flux has decayed to something
of order $\exp(-\omega/T_H)\sim 10^{-20}$!

For the metrics with ``kinks", i.e. regions of large curvature localized
along a static timelike worldline, the agreement with the thermal 
prediction is still remarkably good at $\omega=T_H$, where the relative
deviation in the flux is of order $T_H/k_0$. As the frequency is 
raised however, the thermal flux drops while the flux from scattering
remains of the same order, so it quickly dominates and becomes many
orders of magnitude larger than the thermal component. This non-thermal
flux amounted to roughly 10\% of the total luminosity for the kinkier 
metrics. The flux exhibited oscillations as a function of frequency which
can be explained by interference between the various scattered
contributions to the flux. At low frequencies the thermal component
also interferes.

Although one does not expect kinks in the smooth classical
background metrics ordinarily adopted in black hole radiation studies,
they might be an important source of particle production
once the back reaction is taken into account,
or perhaps in the early universe. It will be interesting to pursue further
the physics of this new sort of particle creation and its possible 
applications.

Even small deviations from a thermal spectrum of black hole radiation
would apparently\cite{gsl} allow violations of the generalized second law 
(GSL), which says that the black hole entropy plus 
the entropy of the exterior cannot decrease. Do the deviations we have found jeapordize the GSL?
Here it is important to recall that even the usual Hawking spectrum is
not that of a blackbody, due to the different absorption coefficients 
for different modes. However this is perfectly consistent with thermal
equilibrium and the generalized second law, since these same absorption
coefficients govern what can be put into a black hole. Similarly, in the
presence of interactions, the Hawking radiation spectrum is shaped by 
those particular interactions, but presumably not in such a way that
allows violations of the GSL. The arguments\cite{GiPe} based on periodicity
on the Euclidean section support this view, as does a perturbative analysis
of a particular interacting theory\cite{LeUn}. For the models considered in 
this paper, it is relevant to point out that the mode conversion 
and scattering processes
underlying the particle creation would also affect what radiation could be 
injected into a black hole. Thus the deviations from thermality are not
necessarily in contradiction with the GSL. It would be worthwile to 
analyze this issue thoroughly, to understand the status of thermal
equilibrium and the GSL in these models.

Finally, the elusive goal of fully understanding the Hawking effect in a theory with a true cutoff on short distance degrees of freedom remains beyond our grasp, although the considerations of \cite{outgoing} seem to provide a 
step in the right direction.

\section{Acknowledgments}
We would like to thank T. Antonsen, R. Brout, S. Massar, M. Ortiz, R. Parentani,
F. Skiff, Ph. Spindel, and W.G. Unruh for helpful discussions. This work was 
supported in part by NSF Grant PHY94-13253, a UMd graduate fellowship, 
and the University of Utrecht.

\appendix
\section{Appendix}
\label{appendix}
We discuss here the exact solution to the ODE (\ref{ode})
obtained for the special case of the kinked free-fall velocity 
\begin{eqnarray}
v(x) = \left\{ \begin{array}{ll}
v_{o}  & x>0 \\
\kappa x + v_{o} & x \leq 0
\end{array}
\right. 
\label{vk}
\end{eqnarray}
where $v_{o}$ and $\kappa$ are negative and positive constants respectively.
Considering first the equation (\ref{ode}), 
with linear velocity $v(x) = (\kappa x + v_{o})$ for {\it all} $x$, we Fourier transform and get a momentum space equation. By defining a new function $\tilde{g}(k)$
via $\tilde{f}(k) = k^{-1 - i \omega/\kappa} e^{-i (v_{o}/\kappa)k} \tilde{g}(k)$, we find that $\tilde{g}(k)$ satisfies the simple looking equation $(d^{2}/dk^{2} + (1 - (k/ k_{o})^{2})/\kappa^{2})\tilde{g}(k)=0$.  The general solution is given by parabolic cylinder functions; however, because we need an $x$ - space solution, we must Fourier transform $\tilde{f}(k)$ which we only know how to do under a more restrictive condition. This condition is $k_{o}/\kappa = 2n+1$ where $n$ is an odd integer. In this case the solutions simplify to solutions of the simple harmonic oscillator problem in quantum mechanics, i.e., $\tilde{g}(k) \propto e^{-k^{2}/(2\kappa k_{o})} H_{n}(\sqrt{\frac{1}{\kappa k_{o}}}k)$.  Inverse Fourier transforming $\tilde{f}(k)$ we get
\begin{eqnarray}
\frac{d^{m}f_{n}(x)}{dx^{m}} & = & (-\sqrt{2n+1} \,\kappa)^{m} \sum_{s=0}^{[n/2]} ((1/4)^{s} \frac{n!}{s!(n-2s)!} e^{-\frac{2n+1}{4} v(x)^{2}} \nonumber \\ 
&   & D_{-i \omega/\kappa +(n-2s-1)+m}(\sqrt{2n+1}\,v(x)))
\label{exsoln}
\end{eqnarray}
where we have included arbitrary order derivatives of $f_{n}$ in this expression since we shall need them later.  $D_{p}(z)$ is the parabolic cylinder function of order $p$.

Going back to the kinked velocity (\ref{vk}), 
since the velocity is constant for positive $x$, the general solution is a sum of four modes, $\sum_{l=1}^{4} c_{l} e^{i k_{l} x}$.  By matching $f_{n}$ given above and its first three derivatives at $x=0$ to this sum, we may solve for the coefficients $c_{l}$ which, as described in the paper, yield the particle production.

At this point the reader may be wondering what happened to the other three linearly independent solutions to the original ODE (\ref{ode}). 
The answer is that we selected one of the solutions by the choice of integration
contour in evaluating the Fourier transform. Moreover, we got lucky, 
because the solution we obtained is the ``right" one: as argued 
in section \ref{sec:ode}, the solution appropriate for
a wavepacket that is purely outgoing at late times should decay across the horizon.  Noting that $D_{p}(z) \rightarrow e^{-z^{2}/4} z^{p}$ as $z \rightarrow \infty$ we see that the above solution satisfies this property. (We did not 
really need this luck since, as explained in section \ref{sec:ode},
almost all the solutions agree outside the horizon up to exponentially 
small terms.) 

Unfortunately, because the exact solution is rather unwieldy, we 
have so far not been able to do anything with it except to use it as a check on our numerical solutions.  
This we did by comparing the predicted particle creation at a given frequency
using the exact solution to that obtained using the numerically generated solution. We found the same results to the numerical accuracy of the 
latter solution, thus confirming the accuracy of the finite differenced
solution. 

\end{document}